\UseRawInputEncoding 
\documentclass[superscriptaddress,aps,prl,reprint,twocolumn,amsmath,amssymb,showpacs,]{revtex4-2}
\usepackage{graphicx}  
\usepackage{mathptmx}
\usepackage{epstopdf}
\usepackage{dcolumn}   
\usepackage{bm}        
\usepackage{amssymb}   
\usepackage{amsmath}   
\usepackage{amsthm}
\usepackage{chemarrow}
\usepackage{color}
\usepackage{mathrsfs}
\usepackage{float}
\usepackage{physics}
\usepackage{bbold}
\usepackage{bbm}
\usepackage[a4paper,colorlinks=true,
linkcolor=blue,citecolor=blue,
pdfauthor={ },
pdftitle={ },
pdfsubject={ },
pdfkeywords={ }]{hyperref}

\theoremstyle{plain}
\newtheorem*{theorem*}{Theorem}

\begin{document}


\title{Floquet Spin Amplification}

\date{\today}

\author{Min Jiang}
\email[]{These authors contributed equally to this work}
\affiliation{
Hefei National Laboratory for Physical Sciences at the Microscale and Department of Modern Physics, University of Science and Technology of China, Hefei 230026, China}
\affiliation{
CAS Key Laboratory of Microscale Magnetic Resonance, University of Science and Technology of China, Hefei 230026, China}
\affiliation{
Synergetic Innovation Center of Quantum Information and Quantum Physics, University of Science and Technology of China, Hefei 230026, China}

\author{Yushu Qin}
\email[]{These authors contributed equally to this work}
\affiliation{
Hefei National Laboratory for Physical Sciences at the Microscale and Department of Modern Physics, University of Science and Technology of China, Hefei 230026, China}
\affiliation{
CAS Key Laboratory of Microscale Magnetic Resonance, University of Science and Technology of China, Hefei 230026, China}
\affiliation{
Synergetic Innovation Center of Quantum Information and Quantum Physics, University of Science and Technology of China, Hefei 230026, China}

\author{Xin Wang}
\email[]{These authors contributed equally to this work}
\affiliation{
Hefei National Laboratory for Physical Sciences at the Microscale and Department of Modern Physics, University of Science and Technology of China, Hefei 230026, China}
\affiliation{
CAS Key Laboratory of Microscale Magnetic Resonance, University of Science and Technology of China, Hefei 230026, China}
\affiliation{
Synergetic Innovation Center of Quantum Information and Quantum Physics, University of Science and Technology of China, Hefei 230026, China}

\author{\mbox{Yuanhong Wang}}
\affiliation{
Hefei National Laboratory for Physical Sciences at the Microscale and Department of Modern Physics, University of Science and Technology of China, Hefei 230026, China}
\affiliation{
CAS Key Laboratory of Microscale Magnetic Resonance, University of Science and Technology of China, Hefei 230026, China}
\affiliation{
Synergetic Innovation Center of Quantum Information and Quantum Physics, University of Science and Technology of China, Hefei 230026, China}

\author{Haowen Su}
\affiliation{
Hefei National Laboratory for Physical Sciences at the Microscale and Department of Modern Physics, University of Science and Technology of China, Hefei 230026, China}
\affiliation{
CAS Key Laboratory of Microscale Magnetic Resonance, University of Science and Technology of China, Hefei 230026, China}
\affiliation{
Synergetic Innovation Center of Quantum Information and Quantum Physics, University of Science and Technology of China, Hefei 230026, China}

\author{Xinhua Peng}
\email[]{xhpeng@ustc.edu.cn}
\affiliation{
Hefei National Laboratory for Physical Sciences at the Microscale and Department of Modern Physics, University of Science and Technology of China, Hefei 230026, China}
\affiliation{
CAS Key Laboratory of Microscale Magnetic Resonance, University of Science and Technology of China, Hefei 230026, China}
\affiliation{
Synergetic Innovation Center of Quantum Information and Quantum Physics, University of Science and Technology of China, Hefei 230026, China}

\author{Dmitry Budker}
\affiliation{Helmholtz-Institut, GSI Helmholtzzentrum f{\"u}r Schwerionenforschung, Mainz 55128, Germany}
\affiliation{Johannes Gutenberg University, Mainz 55128, Germany}
\affiliation{Department of Physics, University of California, Berkeley, CA 94720-7300, USA}

\begin{abstract}
Detection of weak electromagnetic waves and hypothetical particles aided by quantum amplification is important for fundamental physics and applications.
However, demonstrations of quantum amplification are still limited;
in particular, the physics of quantum amplification is not fully explored in periodically driven (Floquet) systems,
which are generally defined by time-periodic Hamiltonians and enable observation of many exotic quantum phenomena such as time crystals.
Here we investigate the magnetic-field signal amplification by periodically driven $^{129}$Xe spins and observe signal amplification at frequencies of transitions between Floquet spin states.
This ``Floquet amplification'' allows to simultaneously enhance and measure multiple magnetic fields with at least one order of magnitude improvement,
offering the capability of femtotesla-level measurements.
Our findings extend the physics of quantum amplification to Floquet systems and can be generalized to a wide variety of existing amplifiers,
enabling a previously unexplored class of ``Floquet amplifiers''.
\end{abstract}

\maketitle

Quantum amplification that offers the capability of enhancing weak signals is ubiquitous and of central importance to various frontiers of science,
ranging from low-noise masers~\cite{breeze2018continuous,jin2015proposal,oxborrow2012room,chu2004electron}, ultra-sensitive magnetic resonance spectroscopy~\cite{suefke2017hydrogen}, weak field and force measurements~\cite{kotler2011single,burd2019quantum,boss2017quantum}, and optical amplifiers~\cite{zavatta2011high} to hypothetical-particle searches beyond the Standard Model~\cite{bradley2003microwave,budker2014proposal,jiang2021search,su2021search}.
To date, the well-established paradigm for realizing signal amplification mostly relies on using inherent discrete transitions of quantum systems~\cite{clerk2010introduction},
including atomic and molecular ensembles~\cite{suefke2017hydrogen,jiang2021search,su2021search,jiang2021floquet}, superconducting qubits, colour centres in diamond~\cite{breeze2018continuous,jin2015proposal,oxborrow2012room}, trapped-ion qubits~\cite{kotler2011single,burd2021quantum}, etc.
However,
realizations of quantum amplification are still limited in practice due to restricted tunability and lack of experimentally accessible frequencies of inherent discrete transitions,
limiting the performance of quantum amplifiers,
\mbox{for example, in operation bandwidth, frequency, and gain.}


Recent years have witnessed an increasing attention to periodically driven (Floquet) systems~\cite{moessner2017equilibration, eckardt2017colloquium},
which can be described by a series of time-independent Floquet states and energies that are analogous to the Brillouin-zone artificial dimension~\cite{shirley1965solution}.
Floquet systems could be a promising platform to explore advanced quantum amplification beyond ordinary systems, partially because such systems provide an ideal way to engineer the inherent discrete states and transitions of quantum systems.
This enables a variety of novel functionalities that might not be otherwise directly accessible,
for example, time crystals~\cite{else2016floquet}, Floquet maser~\cite{jiang2021floquet,liu2021masing},
Floquet Raman transition~\cite{shu2018observation},
prethermalization~\cite{peng2021floquet},
Floquet cavity electromagnonics~\cite{xu2020floquet}, and
Floquet polaritons~\cite{clark2019interacting}.
The potential combination of quantum amplification and Floquet systems
may open opportunities for developing new quantum amplifiers with improved performance,
for example, in operation bandwidth, frequency tunability, and evading the need for population inversion.
Such amplifiers would find promising applications in quantum metrology, for example, simultaneous sensing of multiple magnetic field at different frequencies~\cite{lang2015dynamical},
measurement
of the worldwide magnetic-background noise (including Schumann resonance)~\cite{fraser1975superconducting},
and searches for axionlike dark matter with multiple \mbox{sensitive windows of particle mass~\cite{budker2014proposal,jiang2021search,bradley2003microwave}.}

In this Letter, we report the theoretical and experimental demonstration of quantum amplification on periodically driven spins.
The key ingredient is the use of an ensemble of long-lived hyperpolarized $^{129}$Xe spins as periodically driven system,
which overlaps with optically polarized $^{87}$Rb atoms in the same vapor cell, with Fermi-contact collisions between them.
Unlike conventional quantum amplification that exploits inherent transitions~\cite{breeze2018continuous,jin2015proposal,oxborrow2012room,suefke2017hydrogen,kotler2011single,burd2019quantum,boss2017quantum,chu2004electron,bradley2003microwave,budker2014proposal,jiang2021search,su2021search},
we demonstrate that the driven $^{129}$Xe spins as an amplifier can simultaneously amplify external magnetic fields that oscillate at frequencies of transitions between Floquet states,
by a factor of more than ten.
In addition, we show that the application of certain periodic driving enables spin-amplification gain below one,
providing an ideal way to suppress environmental magnetic noise disturbance.
The present amplification phenomena on Floquet systems are collectively named ``Floquet amplification''.
In contrast to the well-known amplification by stimulated emission of radiation (maser) that requires population inversion~\cite{breeze2018continuous,jin2015proposal,oxborrow2012room,suefke2017hydrogen},
Floquet amplification removes population-inversion requirement and highlights the practical applications of building a new class of quantum amplifiers that could operate in ambient conditions.
As a first application,
our amplifier constitutes a new technology for measuring magnetic fields with broad bandwidth and fT/Hz$^{1/2}$-level sensitivity,
which is notably better than that of other state-of-the-art magnetometers demonstrated
with nuclear spins limited to a sensitivity of a few pT/Hz$^{1/2}$.
The present amplification technique also allows one to search for hypothetical particles with a sensitivity well beyond the most stringent existing constraints~\cite{jiang2021search, garcon2019constraints}.



\begin{figure}[t]  
	\makeatletter
	\def\@captype{figure}
	\makeatother
	\includegraphics[scale=0.7]{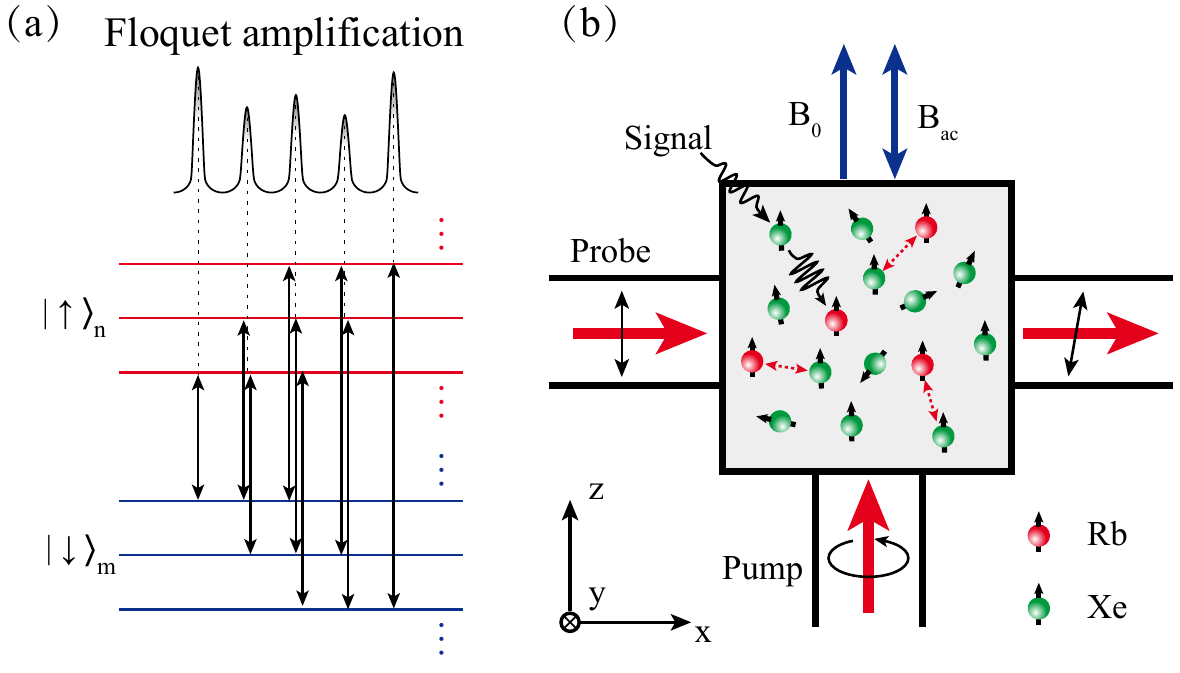}
	\caption{(color online)
	(a) A periodically driven two-level system can be described by a series of time-independent Floquet states $\ket{\uparrow}_n$, $\ket{\downarrow}_m$. We study the transitions between these Floquet states.
    (b) Schematic of experimental setup. The key element is a cubic vapor cell containing 5 torr $^{129}$Xe, 250~torr N$_2$, and a droplet of isotopically enriched $^{87}$Rb. $^{87}$Rb atoms are polarized by a circularly polarized laser at 795~nm and probed by a linearly polarized light blue-detuned 110~GHz from the D2 transition at 780 nm. $^{129}$Xe spins are polarized by spin-exchange collisions with polarized $^{87}$Rb atoms. An oscillating magnetic field as a periodic driving on $^{129}$Xe is applied along $\hat{z}$. The driven $\rm ^{129}Xe $ can amplify the magnetic field that oscillates at frequencies of transitions between Floquet states [see (a)]. The amplified field then is \emph{in situ} read out by $^{87}$Rb magnetometer.
     }
	\label{figure-1}
\end{figure}

We employ a two-level spin system as a testbed of Floquet amplification.
A bias magnetic field $B_{\rm 0}$ is applied along $\hat{z}$,
where the spin Larmor frequency is $\nu_0=\gamma_n B_{\rm 0} $ and $\gamma_n$ denotes the gyromagnetic ratio.
To periodically drive the spins into a Floquet system,
we introduce a radio-frequency (rf) field $B_{\rm ac}{\rm cos}(2\pi\nu_{\rm ac}t) \hat{z}$.
As a result, the spin Hamiltonian $H(t)$ becomes time-periodic but can be mapped into a time-independent Floquet Hamiltonian with infinite dimension~\cite{shirley1965solution, jiang2021floquet}.
As shown in Fig.~\ref{figure-1}(a),
the original two-level system now is expanded to infinite Floquet levels with equal interval $\nu_{\rm ac}$,
i.e.,
$ \ket{\uparrow} $ turns into $ \ket{\uparrow}_n$,
and $ \ket{\downarrow} $ turning into $ \ket{\downarrow}_m$~\cite{SI},
where $n$ and $m$ are non-negative integers.
With synthetic dimensions supported by Floquet states,
the number of resonance transitions increases, thus improving operation bandwidth and
enabling to constitute multi-mode amplifiers.

We perform Floquet-amplification experiments with $\rm ^{129}Xe$ noble gas.
The setup is described in the Supplemental materials~\cite{SI} in detail and is similar to that of Refs.~\cite{jiang2021search,su2021search}.
As shown in Fig.~\ref{figure-1}(b),
5-torr $\rm ^{129}Xe$ gas overlaps with enriched $\rm ^{87}Rb$ in a 0.5-$\rm cm^{3}$ cubic vapor cell.
The vapor cell is enclosed in a boron nitride ceramics and heated to $\sim$160$^{\circ}$C with twisted-pair copper wires.
Circularly polarized laser light tuned to D1 line at $795$ $\rm nm$ polarizes $\rm ^{87}Rb$ atoms, and
$\rm ^{129}Xe$ spins are polarized through spin-exchange collisions with polarized $\rm ^{87}Rb$ atoms.
The spin polarization of $\rm ^{129}Xe$ gas is estimated to be about $30\%$.
We introduce a rf field $B_{\rm{ac}}$ along $\hat z$ for periodic driving of $\rm ^{129}Xe$.
When a transverse oscillating magnetic field matches one of Floquet $\rm ^{129}Xe$ transitions [for example, $\ket{\downarrow}_m \rightarrow  \ket{\uparrow}_n$ in Fig.~\ref{figure-1}(a)],
the polarized $\rm ^{129}Xe$ spins are tilted away from the $\hat{z}$ direction and accordingly generate oscillating transverse magnetization $\textbf{M}_{n}(t)$.
The Fermi-contact collisions between $^{129}$Xe and $^{87}$Rb lead to that $\rm ^{87}Rb$ experiences an effective magnetic field~\cite{bulatowicz2013laboratory,sheng2014new,li2016rotation},
$\textbf{B}_{\textrm{eff}}= (8 \pi \kappa_{0}/3) \textbf{M}_n (t)$,
which is \emph{in situ} read out by $^{87}$Rb magnetometer~\cite{budker2007optical,jiang2020interference,jiang2019magnetic}.
Here $\kappa_0 \approx 540$ denotes the Fermi-contact enhancement factor between $^{129}$Xe and $^{87}$Rb~\cite{walker1997spin}.
As a result, the magnetic field $\mathbf{B}_{\rm{eff}}$ generated by $^{129}$Xe nuclear magnetization can be enhanced by a large factor of $\kappa_0$, and can be significantly larger than the measured field.

Using synthetic dimensions supported by Floquet states,
the magnetic field can be amplified at a series of comb-like frequencies $\nu_0,\nu_0\pm \nu_{\rm{ac}},\cdots, \nu_0\pm k\nu_{\rm{ac}},\cdots$,
which correspond to $k$-photon transitions (or sidebands).
We show that the $k$-photon amplification to the signal is~\cite{SI}
\begin{equation}
    \eta_{k,0}(u)=\frac{4 \pi}{3} \kappa_0 M^n P_0^n\gamma_n T_{2n}{J}_k^2(u),
    \label{eq1}
\end{equation}
where $P_0^n$ is the equilibrium polarization of $\rm ^{129}Xe$,
$M^n$ is the magnetization $\rm ^{129}Xe$ atoms with unity polarization, $\gamma_n\approx0.01178$ Hz/nT is the gyromagnetic ratio of $\rm ^{129}Xe$,
$T_{2n}\approx 34$ s is the transverse relaxation time of $\rm ^{129}Xe$ spins,
${J}_k$ is the Bessel function of the first kind of order $ k $,
$ u = \gamma B_{\rm ac}/ \nu_{\rm ac} $ is modulation index.


\begin{figure}[t]  
	\makeatletter
	\def\@captype{figure}
	\makeatother
	\includegraphics[scale=1.0]{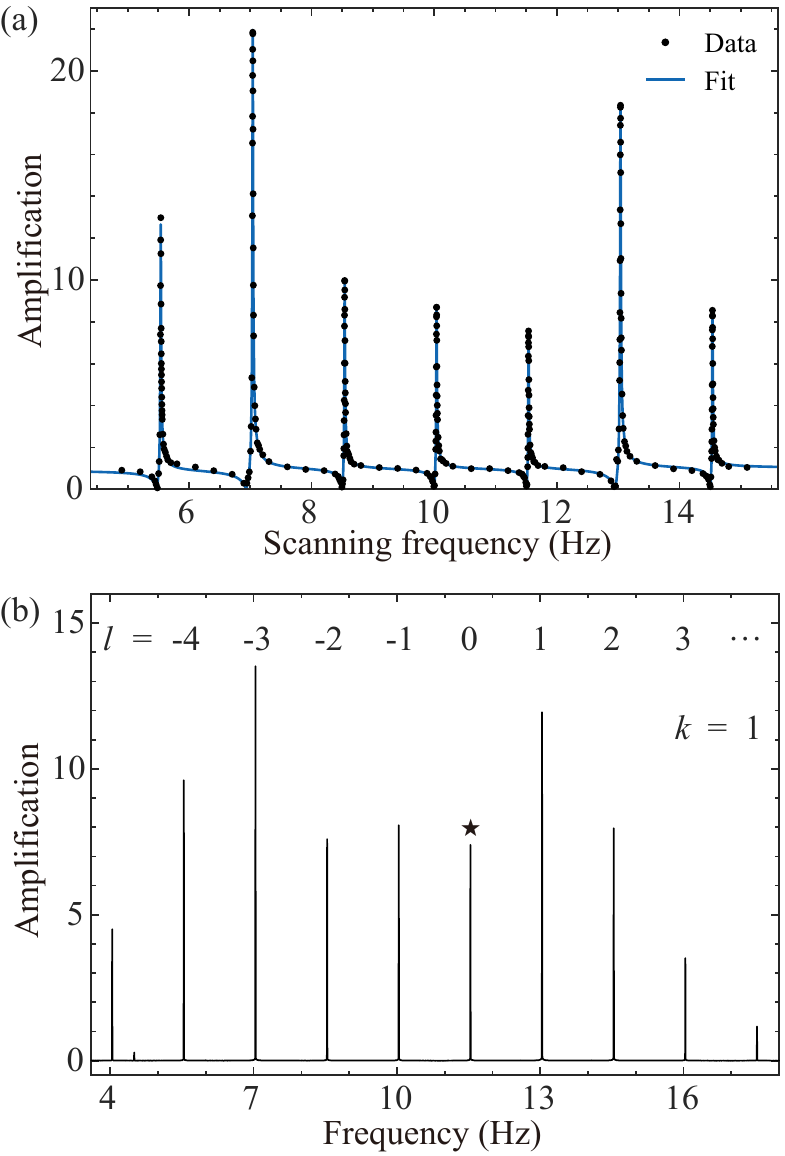}
	\caption{(color online)
	(a) Plot of amplification as a function of the scanning test-field frequency. The bias field and periodic driving field are set as $B_{0}\approx853$ nT, $B_{\rm ac}\approx397$ nT, $\nu_{ \rm{ac}}\approx 1.500$ Hz. The corresponding modulation index is $u \approx 3.12$. The interval between amplification frequencies is 1.500 Hz that is equal to $\nu_{\rm{ac}}$. The amplification at each resonance frequency satisfies $\eta_{k,0}$ (see Eq.~\ref{eq1}). The amplification profile is asymmetric due to the Fano interference described in the text. (b) Spectrum of the amplification signal induced by a test field. The test signal is set at the frequency of 1st-order sideband (see star) and there simultaneously exists other sidebands signals. The amplification satisfies $\eta_{k,l}$ (see Eq.~\ref{eq2}).}
	\label{figure-2}
\end{figure}

Figure~\ref{figure-2}(a) shows the experimental data on Floquet amplification.
The parameters are set as $B_{0}\approx853$ nT, \mbox{$B_{\rm ac}\approx397$ nT}, $\nu_{\rm{ac}}\approx 1.500$ Hz,
corresponding to $\nu_0 \approx 10.039$~Hz and modulation index $ u \approx 3.12$.
A transverse oscillating magnetic field as a test signal is applied along $\hat{y}$, and its frequency is scanned from about 4.900 Hz to 15.100 Hz.
Considering that the resonance linewidth is as narrow as $\approx$17~mHz determined by $\sqrt{3}/(\pi T_{2n})$~\cite{SI},
the scanning frequency step is set as 1 mHz around resonance frequencies.
To determine the amplification factor,
we record $^{87}$Rb magnetometer response signal far way from resonance as a reference.
All signal strengths are obtained by performing Fourier transform of the magnetometer signals.
The experimental result is shown in Fig.~\ref{figure-2}(a) and the ratio among different transitions is close to the theoretical value \mbox{${J}_0^2(u):{J}_{\pm 1}^2(u):{J}_{\pm 2}^2(u):{J}_{\pm 3}^2(u)=1:0.969:2.660:1.225$.}


An amplification phenomenon is observed on Floquet $^{129}$Xe spins.
Specifically,
although the test frequency matches only one Floquet transitions,
there simultaneously exist amplification signals at other Floquet transitions.
For example shown in Fig.~\ref{figure-2}(b),
there is a peak at the frequency of the test field set at 11.539 Hz (marked with a star),
and otherwise there exist other sidebands even with larger signal amplitude.
These unexpected sidebands can be seen as the result of the test-field photons scattered by driven $^{129}$Xe spins,
where the virtual absorption and emission of photons occur~\cite{SI}.
We obtain the corresponding amplification factor~\cite{SI}
\begin{equation}
    \eta_{k,l}(u)=\frac{4}{3}\kappa_0 M^n P_0^n\gamma_n T_{2n}{J}_{l+k}(u){J}_{k}(u),
    \label{eq2}
\end{equation}
where $ k $ is the order of the sideband where the test field is located,
$ l $ is the order difference between the test-field sideband and other induced sideband [see an example in Fig.~\ref{figure-2}(b)].
Obviously,
the amplification factor $\eta_{k,l} (u)$ becomes the same with that described in Eq.~\ref{eq1} when $l=0$.
In the experiment of Fig.~\ref{figure-2}(b),
$k=1$ (matching the first-order sideband).
We obtain the experimental ratio between different cross-amplification lines, i.e., $\eta_{1,-1}:\eta_{1,0}:\eta_{1,1}:\eta_{1,2} \approx 1:0.916:1.480:0.987$,
which is in agreement with the theoretical value:$ |{J}_0(u)|:|{J}_1(u)|:|{J}_2(u)|:|{J}_3(u)| = 1:0.984:1.631:1.107 $. 

\begin{figure}[t]  
	\makeatletter
	\def\@captype{figure}
	\makeatother
	\includegraphics[scale=1.0]{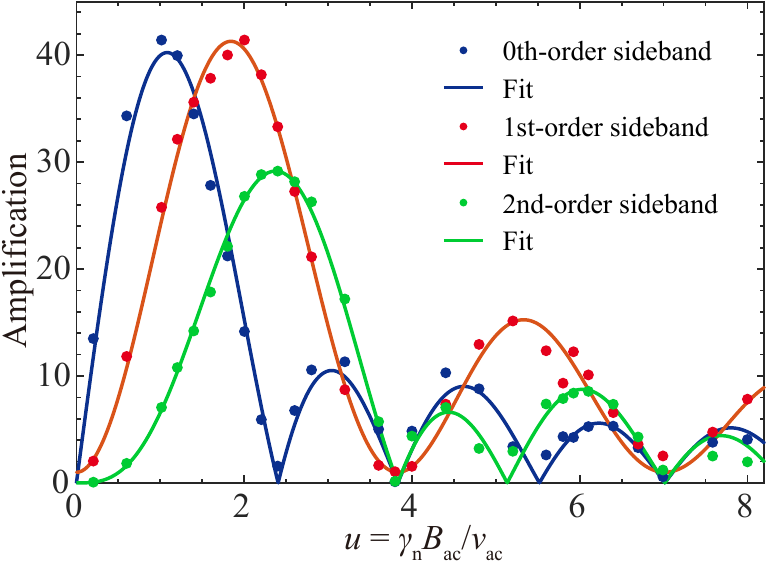}
	\caption{(color online) Plot of Floquet amplification as a function of the modulation index $u$. A bias magnetic field $B_{0} \approx 853$ $\rm nT$ is applied along $\hat z$, corresponding to $\rm ^{129}Xe$ resonance frequency $\nu_0 \approx 10.039$ $\rm Hz$. A periodic driving field of $\nu_{\rm ac} \approx 3.000$ $\rm Hz$ is introduced. By scanning the amplitude $B_{\rm ac}$, the amplification of the test signal at $7.039$ $\rm Hz$ (red points) follows $J_{1}^{2}(u)$ (red line). The amplification at $10.039$ $\rm Hz$ (blue points) follows $J_{1}(u)J_{0}(u)$ (blue line) and at $4.039$ $\rm Hz$ (green points) follows $J_{1}(u)J_{2}(u)$ (green line).}
	\label{figure-3}
\end{figure}

We find an analytical relation for power amplification:
\begin{equation}
\sum_{l=-\infty}^{+\infty}\eta_{k,l}^2(u)=\eta_0(0)\eta_k(u),
\end{equation}
indicating the total power at all Floquet transitions remaining a constant for a measurement.
This kind of cross-amplification is well suited for eliminating uncorrelated noise.
For example, we employ seven sidebands shown in Fig.~\ref{figure-2}(b) as detection indicators.
A true event should induce detectable signals at all sidebands.
In contrast to the detection with a single sideband where
the noise-induced false alarm rate is $ 5\% $ ($95\%$ confidence level),
the simultaneous detection with seven sidebands would reduce the false-positive rate down to $7.8\times 10^{-10}$,
with seven orders of magnitude improvement.
This would provide a high-confidence-level way to identify an event from random noise.

The amplification $\eta_{k,l}(u)$ is adjustable by changing modulation index $u$ according to Eq.~\ref{eq2}.
We set the resonance frequency of $\rm ^{129}Xe$ at $\nu_{0} \approx 10.039$ $\rm Hz$ and the periodic driving frequency at $\nu_{\rm{ac}} \approx 3.000$ $\rm Hz$.
The index $u$ is scanned from 0 to 8.00 by changing driving amplitudes.
The test-field frequency is applied, for example, at the first-order sideband $\nu_0 - \nu_{\rm{ac}} \approx 7.039$ $\rm Hz$.
In this case, we plot the amplification factors of three sidebands as a function of $u$, as shown in Fig.~\ref{figure-3}.
The zeroth at $\nu_0$, first-sideband at $\nu_0-\nu_{\rm{ac}}$ and second-sideband at $\nu_0-2\nu_{\rm{ac}}$ well follows theoretical profiles $J_{1}(u)J_{0}(u)$ (blue line), $J_{1}^{2}(u)$ (red line) and $J_{1} (u)J_{2}(u)$ (green line), respectively.
The modulation index $u$ can be optimized to maximize the amplification.
For example, when $u$ is close to the theoretical value 1.84,
the first-order amplification (red line) reaches the maximum $41.4$ (i.e., 32.3~dB).

The line shape of amplification as a function of scanning frequency is not symmetric under certain regime, as shown in Fig.~\ref{figure-2}(a).
We find that the asymmetric line originates from the well-known Fano resonance~\cite{luk2010fano,limonov2017fano}.
Specifically,
the $\rm ^{87}Rb$ and $\rm ^{129}Xe$ spins simultaneously experience the applied test field and both independently generate signals at the test frequency.
To appear the Fano resonance,
there are a discrete system and a continuum system.
In our case, $\rm ^{129}Xe$ spins have sharp resonance line and thus can be seen as a discrete system,
whereas $\rm ^{87}Rb$ spins have broad resonance line and can be approximated as a continuum system.
The phase of $\rm ^{129}Xe$ induced signal changes rapidly near resonance,
while the phase of $\rm ^{87}Rb$ signal varies slowly.
Due to the phase difference between their signals,
the $\rm ^{129}Xe$ induced and direct $\rm ^{87}Rb$ signals interfere with each other and their interference gives rise to the asymmetric line shape,
where a destructive interference occurs on the left and a constructive interference occurs on the right side of the resonance [see Fig.~\ref{figure-2}(a)].
Notably, due to the destructive interference,
the amplification factor could be smaller than one when the scanning test frequency is nearby Floquet transitions.
A detailed theoretical model of Fano resonance is presented in the Supplemental materials~\cite{SI}.

\begin{figure}[t]  
	\makeatletter
	\def\@captype{figure}
	\makeatother
	\includegraphics[scale=1.0]{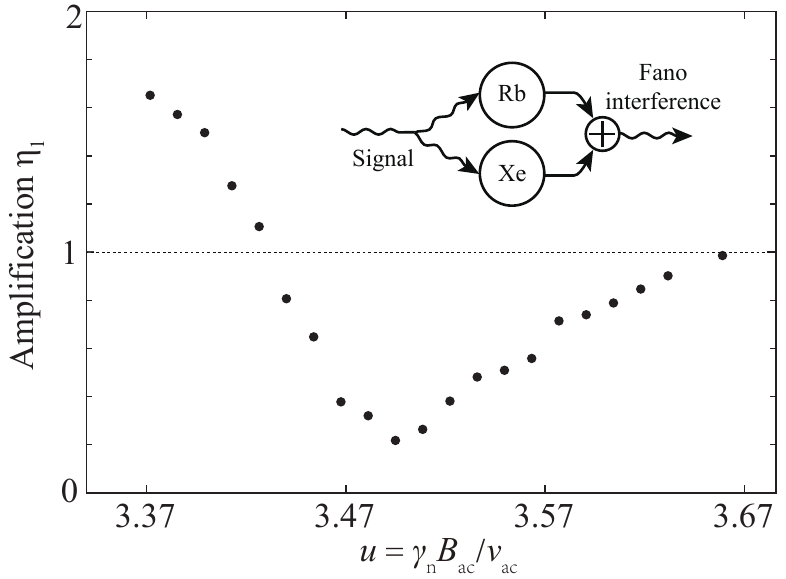}
	\caption{(color online) Fano resonance in spin amplification. Black points denote the measured amplification of the test-field signal at $2.970$ $\rm Hz$ with the parameters of $\nu_{0} \approx 10.039$ $\rm Hz$ and $\nu_{\rm ac} \approx 13.000$ $\rm Hz$. The signal amplitude is suppressed by a factor of 5 when the modulation index is $u \approx 3.497$. In the inset, the $^{129}$Xe-$^{87}$Rb signals interfere with each other and their interference results in suppressing amplification.}
	\label{figure-4}
\end{figure}

Because of the existence of Fano resonance described above,
the amplifier reduces the response to the test field and can be adjusted to be insensitive to environmental magnetic noise.
We quantify the response reduction under different modulation index $u$.
We set $\nu_{0} \approx 10.039$ $\rm Hz$ and $\nu_{\rm ac} \approx 13.000$ $\rm Hz$.
The first-order sideband frequency is at $2.971$ $\rm Hz$.
When the modulation index $u$ is scanned from 3.44 to 3.66 by changing $B_{\rm ac}$,
the amplification factor is smaller than one,
turning into de-amplification regime.
For example, the amplification reduces down to $\eta_{1} \approx 0.2$ when $u\approx 3.50$,
indicating that the response of magnetic field noise can be suppressed by a factor of about five.
As shown in the inset of Fig.~\ref{figure-4},
when the effective field of $\rm ^{129}Xe$ and test signal have different phase,
destructive interference occurs and noise suppression appears.
Similar mechanism of noise suppression has been reported in self-compensating comagnetometer~\cite{kornack2005nuclear,li2016rotation}.
However, comagnetometers usually work for suppressing adiabatically changing (low-frequency) magnetic noise,
our amplifier provides the capability of suppressing high-frequency magnetic noise.
Although insensitive to actual magnetic field,
our amplifier remains sensitive to some beyond-the-Standard-Model exotic fields that couple with only one of $^{129}$Xe and $^{87}$Rb~\cite{jiang2021search,su2021search,safronova2018search,bradley2003microwave}.

The present Floquet amplification can be used for a broad range of applications in precision measurements.
For example,
the application of Floquet amplification to magnetic field sensing,
which is capable of simultaneously measuring fields at Floquet transitions,
yields one order of magnitude improvements over previously achievable detection bandwidth~\cite{jiang2021search, su2021search}.
Based on the demonstrations in this work,
the magnetic field sensitivity is in the range of tens fT/Hz$^{1/2}$, and can be further improved to aT/Hz$^{1/2}$-level using K-$^{3}$He system due to the longer $^3$He coherence time.
Our amplifier can be applied
to search for hypothetical particles predicted by numerous theories beyond the Standard Model~\cite{graham2018spin,graham2015experimental},
such as ultralight axions and dark photons.
These particles are predicted to couple with Standard-Model particles (such as nuclear spins) and behave as an oscillating magnetic field~\cite{budker2014proposal,graham2013new},
which can be greatly amplified with our amplifier.
As a result, it is promising to improve the search sensitivity
of axions and dark photons with new limits beyond the astrophysical ones.
Importantly, to complete the searches for a full range of particle masses,
the traditional approaches have to perform measurements and scan with a single bandwidth~\cite{budker2014proposal,graham2013new,jiang2021search,bradley2003microwave,aybas2021search,arvanitaki2014resonantly},
which is time-consuming and limits experimental searches.
In contrast,
our approach can simultaneously measure the exotic fields with multiple detection bandwidth.
Moreover, the Floquet amplification provides an ideal way to distinguish the exotic-field signal from spurious random noise with a false-positive rate as low as $\sim$$10^{-9}$.

In conclusion,
we have reported a demonstration of spin amplification on periodically driven (Floquet) $^{129}$Xe spins.
The successful combination of quantum amplification and Floquet systems allows for observing a class of Floquet amplification phenomena,
which are not accessible in previous studies.
Our findings improve the detection bandwidth by one order of magnitude and make it possible to build a new type of quantum amplifiers that allow simultaneously amplifying the electromagnetic wave at multiple frequencies.
Such properties of amplifiers will be essential for low-energy searches for hypothetical particles beyond the Standard Model~\cite{jiang2021search,su2021search}.
Although demonstrated for $^{129}$Xe system,
our scheme of Floquet amplification is generic and can be applied to a wide range of quantum amplifiers.
For example, recent advances in pentacene molecules~\cite{oxborrow2012room,wu2021bench} and nitrogen-vacancy defect materials~\cite{breeze2018continuous,jin2015proposal} have led to progress in areas of low-noise room-temperature amplifiers.
The combination of such amplifiers and periodic driving (for example, with the use of an oscillating magnetic field) could increase the detection regimes that could find attractive applications in cosmological observation and deep-space communications.

~\

We thank Dong Sheng, Hao Wu, Jianmin Cai, and \mbox{Renbao Liu} for valuable discussions.
This work was supported by National Key Research and Development Program of China (grant no.~2018YFA0306600), National Natural Science Foundation of China (grants nos. 11661161018, 11927811, 12004371), Anhui Initiative in Quantum Information Technologies (grant no.~AHY050000), and USTC Research Funds of the Double First-Class Initiative (grant no. YD3540002002).
This work was also supported in part by the Cluster of Excellence ``Precision Physics, Fundamental Interactions, and Structure of Matter'' (PRISMA+ EXC 2118/1) funded by the German Research Foundation (DFG) within the German Excellence Strategy (Project ID 39083149).

\bibliographystyle{naturemag}
\bibliography{mainrefs}

\end{document}


\title{Supplemental materials for ``Floquet Spin Amplification"}

\date{\today}

\author{Min Jiang}
\email[]{These authors contributed equally to this work}
\affiliation{
Hefei National Laboratory for Physical Sciences at the Microscale and Department of Modern Physics, University of Science and Technology of China, Hefei 230026, China}
\affiliation{
CAS Key Laboratory of Microscale Magnetic Resonance, University of Science and Technology of China, Hefei 230026, China}
\affiliation{
Synergetic Innovation Center of Quantum Information and Quantum Physics, University of Science and Technology of China, Hefei 230026, China}

\author{Yushu Qin}
\email[]{These authors contributed equally to this work}
\affiliation{
Hefei National Laboratory for Physical Sciences at the Microscale and Department of Modern Physics, University of Science and Technology of China, Hefei 230026, China}
\affiliation{
CAS Key Laboratory of Microscale Magnetic Resonance, University of Science and Technology of China, Hefei 230026, China}
\affiliation{
Synergetic Innovation Center of Quantum Information and Quantum Physics, University of Science and Technology of China, Hefei 230026, China}

\author{Xin Wang}
\email[]{These authors contributed equally to this work}
\affiliation{
Hefei National Laboratory for Physical Sciences at the Microscale and Department of Modern Physics, University of Science and Technology of China, Hefei 230026, China}
\affiliation{
CAS Key Laboratory of Microscale Magnetic Resonance, University of Science and Technology of China, Hefei 230026, China}
\affiliation{
Synergetic Innovation Center of Quantum Information and Quantum Physics, University of Science and Technology of China, Hefei 230026, China}

\author{\mbox{Yuanhong Wang}}
\affiliation{
Hefei National Laboratory for Physical Sciences at the Microscale and Department of Modern Physics, University of Science and Technology of China, Hefei 230026, China}
\affiliation{
CAS Key Laboratory of Microscale Magnetic Resonance, University of Science and Technology of China, Hefei 230026, China}
\affiliation{
Synergetic Innovation Center of Quantum Information and Quantum Physics, University of Science and Technology of China, Hefei 230026, China}

\author{Haowen Su}
\affiliation{
Hefei National Laboratory for Physical Sciences at the Microscale and Department of Modern Physics, University of Science and Technology of China, Hefei 230026, China}
\affiliation{
CAS Key Laboratory of Microscale Magnetic Resonance, University of Science and Technology of China, Hefei 230026, China}
\affiliation{
Synergetic Innovation Center of Quantum Information and Quantum Physics, University of Science and Technology of China, Hefei 230026, China}

\author{Xinhua Peng}
\email[]{xhpeng@ustc.edu.cn}
\affiliation{
Hefei National Laboratory for Physical Sciences at the Microscale and Department of Modern Physics, University of Science and Technology of China, Hefei 230026, China}
\affiliation{
CAS Key Laboratory of Microscale Magnetic Resonance, University of Science and Technology of China, Hefei 230026, China}
\affiliation{
Synergetic Innovation Center of Quantum Information and Quantum Physics, University of Science and Technology of China, Hefei 230026, China}

\author{Dmitry Budker}
\affiliation{Helmholtz-Institut, GSI Helmholtzzentrum f{\"u}r Schwerionenforschung, Mainz 55128, Germany}
\affiliation{Johannes Gutenberg University, Mainz 55128, Germany}
\affiliation{Department of Physics, University of California, Berkeley, CA 94720-7300, USA}

\maketitle

\tableofcontents

~\

\begin{figure}[t]  
	\makeatletter
\centering
	\def\@captype{figure}
	\makeatother
	\includegraphics[scale=0.5]{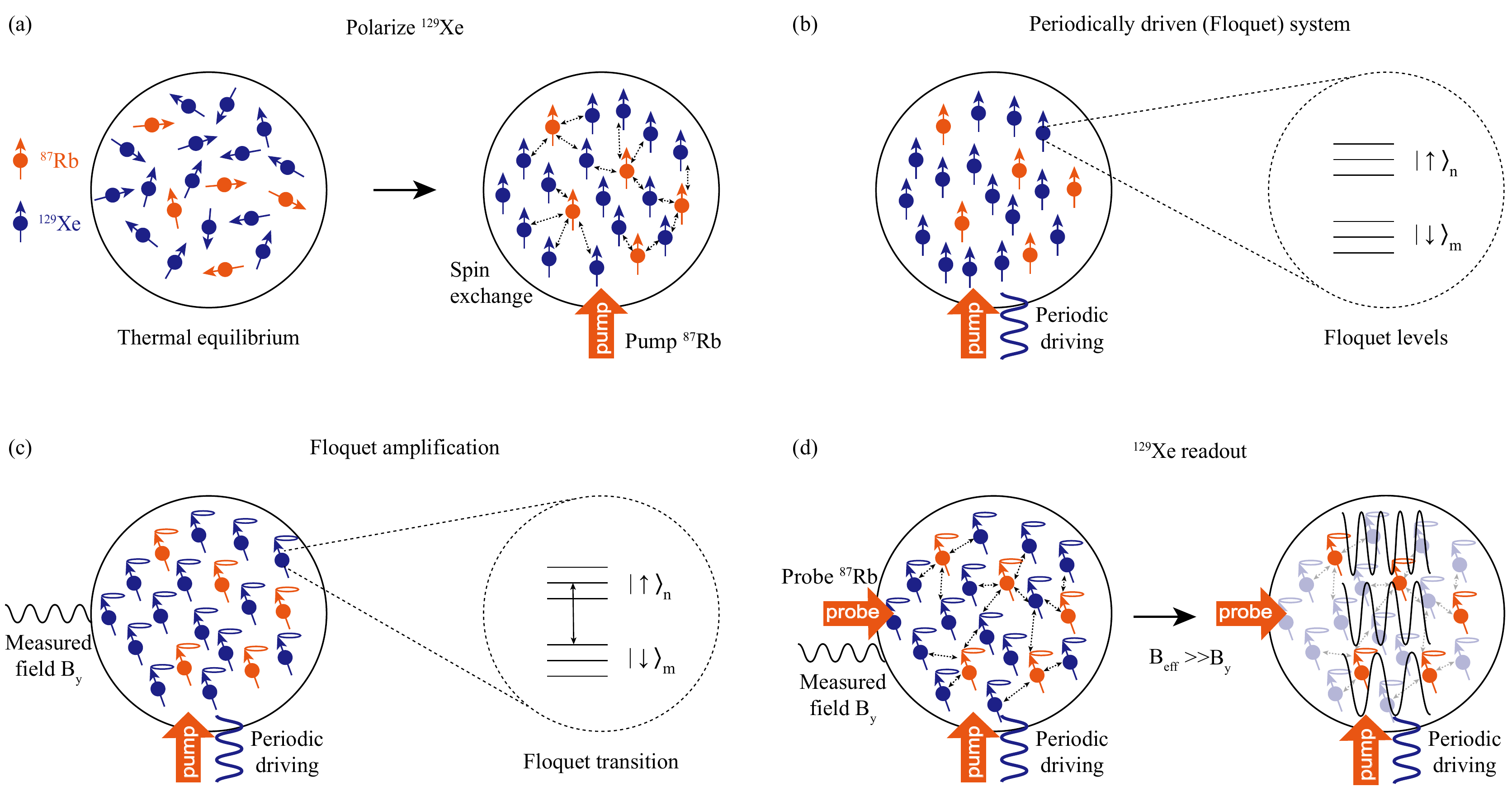}
	\caption{Schematic diagram of Floquet amplification. A 0.5-cm$^3$ cubic vapor cell contains 5~torr of isotopically enriched $\rm ^{129}Xe$, 250~torr N$_2$ as buffer gas, and a droplet (several miligrams) of isotopically enriched $^{87}$Rb metal. (a) Circularly polarized laser light tuned to the D1 line of $^{87}$Rb at 795 nm polarizes the $^{87}$Rb atoms along $\hat{z}$. $\rm ^{129}Xe$ spins are polarized through spin-exchange collisions with optically polarized $\rm ^{87}Rb$ atoms. (b) $^{129}$Xe spins are periodically modulated by an oscillating magnetic field along $\hat{z}$. As a result, two-level $^{129}$Xe spins can be described by a series of time-independent Floquet states and energies. (c) An oscillating transverse field $B_{\rm{y}}$ can be amplified when its frequency matches one of transition frequencies between Floquet states. (d) $\rm ^{129}Xe$ spins amplify the external measured field and generate an effective field on $\rm ^{87}Rb$ atoms. $\rm ^{87}Rb$ atoms can function as a sensitive magnetometer to \emph{in situ} read out the effective field via the optical rotation of a linearly polarized probe laser along $\hat{x}$.}
	\label{figure1-2}
\end{figure}

\begin{figure}[t]  
	\makeatletter
\centering
	\def\@captype{figure}
	\makeatother
	\includegraphics[scale=1.16]{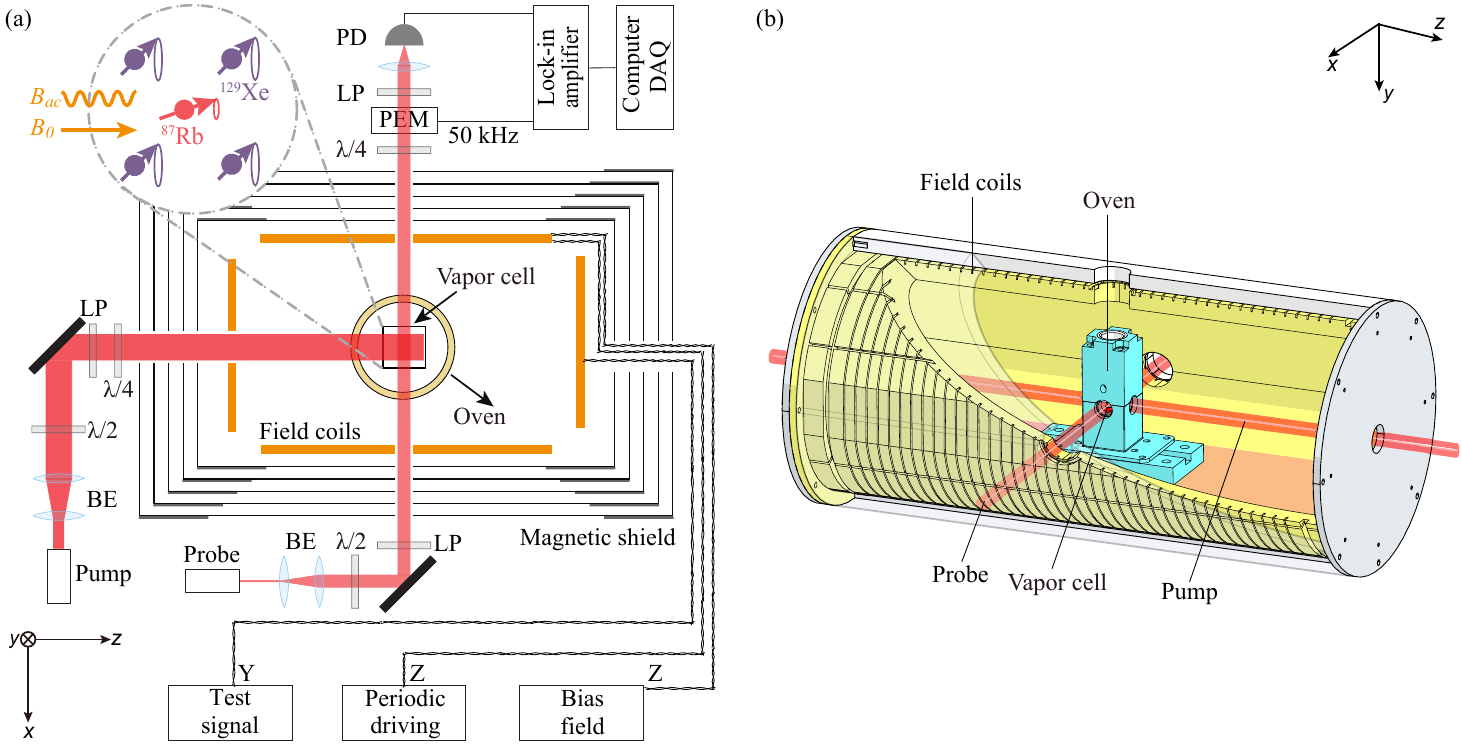}
	\caption{Experimental apparatus. (a) $^{87}$Rb atoms are polarized with a circularly polarized pump laser at 795~nm and probed with a linearly polarized probe laser blue-detuned 110~GHz from the D2 transition at 780 nm. $\rm ^{129}Xe$ spins are polarized and read out through spin-exchange collisions with the polarized $\rm ^{87}Rb$. A bias magnetic field $B_{\rm 0}$ is applied along $\hat{z}$. An oscillating field $B_{\rm{ac}}$ as the periodic driving for tuning $\rm ^{129}Xe$ into a Floquet system are applied along $\hat z$. BE, beam expander; $\lambda/2$, half-wave plate; LP, linearly polarizer; $\lambda/4$, quarter-wave plate; PEM, photoelastic modulator; PD, photo detector; DAQ, data acquisition. (b) Three-dimensional diagram of experimental apparatus. A bias field and a periodic driving field are independently applied with two sets of solenoid coils along $\hat z$. \mbox{The test signal is applied with a saddle coil along $\hat y$.}}
	\label{figure1}
\end{figure}

\section{Overview of Floquet amplification and experimental apparatus}
\label{sec1}

This section presents an overview of Floquet amplification and describes the details of experimental apparatus.
In a Floquet-amplification experiment,
the key element is a 0.5-cm$^3$ cubic vapor cell,
which contains 5\,torr of isotopically enriched $\rm ^{129}Xe$, 250\,torr N$_2$ as buffer gas,
and a droplet (several miligrams) of isotopically enriched $\rm ^{87}Rb$ metal.
The vapor cell is placed inside of a five-layer cylindrical $\mu$-metal shield (shielding factor $\sim$\,$10^6$).
The $^{129}$Xe-$^{87}$Rb vapor cell is enclosed in a boron nitride ceramics and heated to $160$~$^\circ$C with twisted-pair copper wires.
$^{129}$Xe spins function as the amplification medium of measured oscillating magnetic fields and then produce an amplified effective fields on $^{87}$Rb atoms.
$^{87}$Rb atoms then act as a sensitive magnetometer to \emph{in situ} read out the effective field.
We summarize the basic operations of Floquet amplification in Fig.~\ref{figure1-2}:
(a) Polarize $^{129}$Xe spins;
(b) Periodically drive $^{129}$Xe spins;
(c) Floquet amplification;
(d) Read out $^{129}$Xe spins.
It is important to emphasize that the above four operations simultaneously occur although they are separately shown in Fig.~\ref{figure1-2}.
We explain the details as follows.

\begin{itemize}
\item[(a)]
\textbf{Polarize $^{129}$Xe} [see Fig.~\ref{figure1-2}(a)].
A bias field $B_{0}$ ($\approx 853$ $\rm{nT}$) is applied along $\hat z$ with the current supplied by a precision current source (Krohn-Hite Model 523).
Accordingly, the Larmor frequency of $^{129}$Xe spins is about $10$ Hz.
As shown in Fig.~\ref{figure1}(a),
circularly polarized laser light tuned to the D1 line of $\rm ^{87}Rb$ at 795 nm polarizes $\rm ^{87}Rb$ atoms along $\hat z$.
The optical power of the pump laser is $\sim$26~$\textrm{mW}$.
$\rm ^{129}Xe$ spins are polarized along $\hat z$ through spin-exchange collisions with the polarized $\rm ^{87}Rb$ atoms.
The $\rm ^{129}Xe$ spin polarization is estimated to be $\sim$\,$ 30\%$.
\end{itemize}

\begin{itemize}
\item[(b)]
\textbf{Periodic driving of $^{129}$Xe}~[see Fig.~\ref{figure1-2}(b)].
To periodically drive $^{129}$Xe spins,
we apply an oscillating magnetic field $B_{\rm{ac}} \cos (2\pi \nu_{\rm{ac}}t) \hat z$ with a waveform generator (Keysight Model 33210A).
Under the periodic driving,
the original two-level $^{129}$Xe system is expanded to infinite Floquet levels with equal interval $\nu_{\rm ac}$,
i.e.,
$ \ket{\uparrow} $ turns into $ \ket{\uparrow}_n$,
and $ \ket{\downarrow} $ turning into $ \ket{\downarrow}_m$~\cite{jiang2021floquet},
where $n$ and $m$ are non-negative integers.
The detailed forms of Floquet states are presented in Sec.~\ref{secnew2}.
\end{itemize}

\begin{itemize}
\item[(c)]
\textbf{Floquet amplification}~[see Fig.~\ref{figure1-2}(c)].
Once the frequency of the measured transverse field $B_y$ matches one of Floquet frequencies,
the polarized $\rm ^{129}Xe$ spins are tilted away from the $\hat{z}$ direction and accordingly generate oscillating transverse magnetization $\textbf{M}_{n}(t)$.
The Fermi-contact collisions between $^{129}$Xe and $^{87}$Rb lead to $\rm ^{87}Rb$ experiencing an effective magnetic field~\cite{bulatowicz2013laboratory,sheng2014new,li2016rotation},
$\textbf{B}_{\textrm{eff}}= (8 \pi \kappa_{0}/3) \textbf{M}_n (t)$,
which is read out \emph{in situ} by the $^{87}$Rb magnetometer~\cite{budker2007optical,jiang2020interference,jiang2019magnetic}.
Here $\kappa_0 \approx 540$ denotes the Fermi-contact enhancement factor between $^{129}$Xe and $^{87}$Rb~\cite{walker1997spin}.
As a result, the magnetic field $\mathbf{B}_{\rm{eff}}$ generated by $^{129}$Xe nuclear magnetization can be enhanced by a large factor of $\kappa_0$,
and can be notably larger than the measured transverse field amplitude $B_y$.
A detailed amplification factor of the measured magnetic field is presented in Sec.~\ref{sec2}.
\end{itemize}

\begin{itemize}
\item[(d)]
\textbf{Readout of $^{129}$Xe effective field}~[see Fig.~\ref{figure1-2}(d)].
In the same vapor cell with $^{129}$Xe,
$\rm ^{87}Rb$ atoms function as a sensitive magnetometer to \emph{in situ} read out the $^{129}$Xe effective field $\textbf{B}_{\rm{eff}}$.
Specifically,
the effective field is measured via the optical rotation of the linearly polarized probe laser beam along the $\hat x$ direction [see Fig.~\ref{figure1-2}(a) and (b)].
The probe beam is blue-detuned by 110 GHz from the D2 transition of $\rm ^{87}Rb$ at 780\,nm.
The probe laser power is $\sim$\,0.8\,$\textrm{mW}$.
The optical rotation of the probe beam after passing the vapor cell is (see, for example, Refs.~\cite{wu1986optical, opechowski1953magneto, jiang2020interference})
\begin{equation}
\theta= lr_e cfn P_x^e D(v_{\rm pr})/4,
\end{equation}
where $P_x^e$ is the electron spin polarization of $^{87}$Rb atoms along the $\hat x$ direction; its explicit form (which depends on the unknown magnetic fields) is described below, $l\approx 8$~mm is the optical path length, $r_e\approx 2.8 \times 10^{-13}$~cm is the classical radius of the electron, $c$ is the speed of light, $f$ is the oscillator strength (about 1/3 for D1 light and 2/3 for D2 light), $D(v_{\rm pr}) = (v_{\rm pr}-v_{\rm D2})/[(v_{\rm pr}-v_{\rm D2})^2+(\Gamma/2)^2]$, $v_{\rm pr}$ is the frequency of the probe laser light, and $\Gamma$ is the full-width at half-maximum (FWHM) of the optical D2 transition of frequency $v_{\rm D2}$.
To suppress the influence of low-frequency noise of the probe beam,
the polarization of the probe beam is modulated with a photoelastic modulator (PEM) upon to 50\,$\rm{kHz}$.
The signal is demodulated with a lock-in amplifier (SRS Model 830) and acquired with a 24-bit acquisition card (NI model 9239).
The processed signal is lastly sent to the computer and recorded.
\end{itemize}

\section{Floquet dynamics}
\label{secnew2}

This section presents a physical picture to understand the spin dynamics under periodic driving.
In experiment,
we employ an AC field along $\hat{z}$ to periodically drive $^{129}$Xe spins.
Based on the second-quantization of the AC field,
the field can be regarded as photons,
in which we introduce creation and annihilation operators $\hat{a}^\dagger$ and $\hat{a}$ to describe the photons.
A two-level spin system driven by the AC field can be treated as a dressed-spin system~\cite{Cohen_Tannoudji_atomphoton}.
Here, we emphasize that
although we use the second-quantization method to deal with the AC field,
the physical results from the second quantization are equivalent to those with a semi-classical method in Sec.~\ref{sec3}~(see, for example, Ref.~\cite{novikov1978nonlinear}).

\subsection{Floquet states and energies}

We first derive the eigenvalues and eigenstates of the two-level spin system driven with an AC magnetic field.
Similar to the derivation in Ref.~\cite{jiang2021floquet},
we take $\hbar=1$ and the spin Hamiltonian is
\begin{equation}
\hat{H}/2\pi= \nu_{0} I_z + \nu_{\textrm{ac}} \hat{a}^\dag \hat{a}+ \lambda I_z (\hat{a}^\dag+ \hat{a}),
\end{equation}
where $\nu_0=|\gamma B_0|$, $I_z=\sigma_z/2$,
$\lambda$ denotes the coupling strength $\lambda=\gamma B_{\rm{ac}}/(2\sqrt{\bar{N}})$ between an AC photon and spin~\cite{Cohen_Tannoudji_atomphoton},
$\gamma$ is the gyromagnetic ratio of $^{129}\rm{Xe}$, $\sigma_z$ is the Pauli matrix, $\nu_{\rm{ac}}$ and $B_{\rm{ac}}$ are respectively the frequency and amplitude of the AC field,
and $\bar{N}\gg 1$ is the average number of AC photons.
The creation and annihilation operators $\hat{a}^\dagger$ and $\hat{a}$ satisfy $[\hat{a},\hat{a}^\dagger]=1$ and $[\hat{a}^\dagger,\hat{a}^\dagger]=[\hat{a},\hat{a}]=0$.
This Hamiltonian can be written as a direct sum $\hat{H}=\hat{H}_{1}\oplus\hat{H}_{-1}$, where
\begin{equation}
\begin{array}{cccc}
\hat{H}_{\epsilon}/2\pi= \frac{\epsilon}{2} \nu_{0} + \nu_{\textrm{ac}} \hat{a}^\dag \hat{a}+ \frac{\epsilon}{2} \lambda  (\hat{a}^\dag+ \hat{a})
\end{array}
\end{equation}
and $\epsilon=\pm1$.
The union of the eigenvalues and eigenstates of these two subspaces constitutes the eigenvalue and eigenstate of $\hat{H}/2\pi$.
The displacement operator is defined as $\textrm{D}(\xi)=e^{\xi \hat{a}^\dagger -\xi^{\ast} \hat{a}}$.
It has some properties: $\textrm{D}(\xi)\textrm{D}(\xi)^\dagger=1$, $\textrm{D}(\xi)\hat{a}^\dagger \textrm{D}(\xi)^\dagger=\hat{a}^\dagger-\xi^{\ast}$, $\textrm{D}(\xi)\hat{a} \textrm{D}(\xi)^\dagger=\hat{a}-\xi$.
Thus, $\hat{H}_{\epsilon}/2\pi$ can be written as
\begin{equation}
\begin{array}{cccc}
\hat{H}_{\epsilon}/2\pi=\textrm{D}(-\frac{\epsilon \lambda}{2\nu_{\textrm{ac}}}) ( \frac{\epsilon}{2} \nu_{0} + \nu_{\textrm{ac}} \hat{a}^\dag \hat{a}-\frac{\lambda^2}{4\nu_{\textrm{ac}}}) \textrm{D}^{\dagger}(-\frac{\epsilon \lambda}{2\nu_{\textrm{ac}}}).
\end{array}
\end{equation}
We can find the eigenstates 
\begin{equation}
\begin{array}{cccc}
|\epsilon\rangle_n=\textrm{D}(-\frac{\epsilon \lambda}{2\nu_{\textrm{ac}}})|\epsilon,n\rangle
\end{array}
\end{equation}
and the energy 
\begin{equation}
\begin{array}{cccc}
E_{\epsilon,n}/2\pi=\epsilon \nu_0/2+n\nu_{\rm{ac}}.
\end{array}
\end{equation}
These eigenstates and energies are respectively called ``Floquet states'' and ``Floquet energies''.
Now we derive  $|\epsilon\rangle_n$ for $n\approx\bar{N}\gg1$ on the basis of $|\epsilon, n\rangle$.
Under the condition where the average number of photons is large enough,
we can adopt the following approximation
\begin{equation}
\begin{array}{cccc}
\langle n-m|\textrm{D}(-\frac{\epsilon \lambda}{2\nu_{\textrm{ac}}})|n\rangle \approx e^{-\frac{\lambda^2}{8\nu_{\rm{ac}}}} {J}_{-m} (\frac{-\epsilon\lambda}{\nu_{\rm{ac}}} \sqrt{\bar{N}})\approx{J}_{m}(\frac{\epsilon \gamma B_{\textrm{ac}}}{2\nu_{\textrm{ac}}}).
\end{array}
\end{equation}
Then, the Floquet state $|\epsilon\rangle_n$ on the basis of $|\epsilon, n\rangle$ becomes
\begin{equation}
\begin{array}{cccc}
|\epsilon\rangle_n &=&|\epsilon\rangle \sum_m |n-m\rangle \langle n-m|\textrm{D}(-\frac{\epsilon \lambda}{2\nu_{\textrm{ac}}})|n\rangle
= \sum_{n'} {J}_{n-n'} (\frac{\epsilon \gamma B_{\textrm{ac}}}{2\nu_{\textrm{ac}}}) |\epsilon, n'\rangle.
\label{Floquetstate}
\end{array}
\end{equation}
This clearly shows that a Floquet state is a superposition of the particle number representation $|\epsilon, n'\rangle$.
Importantly, the transition between two different Floquet states can be represented by a series of transitions between $|\epsilon, n'\rangle$ (see Fig.~\ref{Floquet_State} for details).

\subsection{Floquet transition}

\begin{figure}[htb]  
	\makeatletter
\centering
	\def\@captype{figure}
	\makeatother
	\includegraphics[scale=1.1]{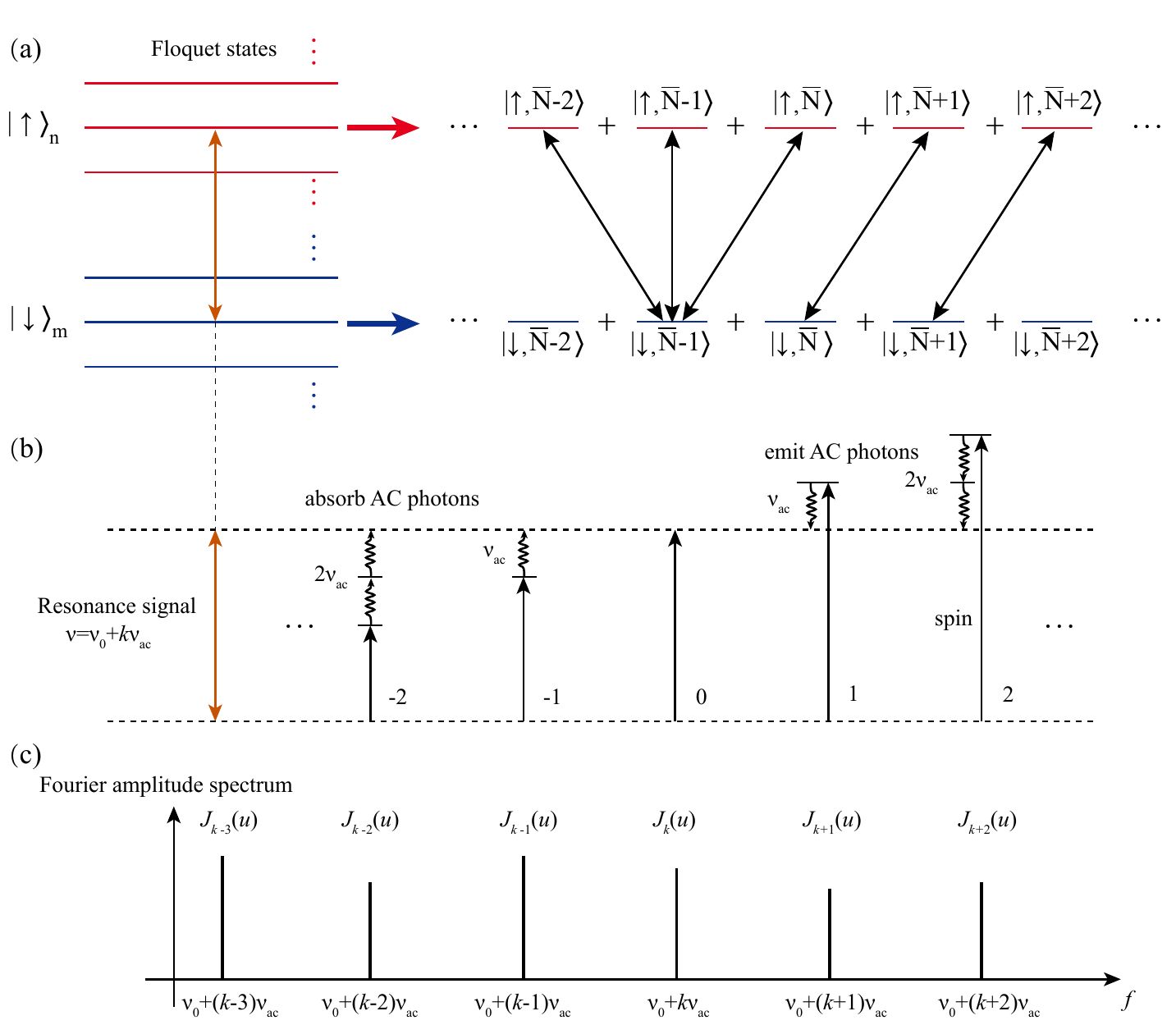}
	\caption{Floquet state and multiple AC photon transitions.}
	\label{Floquet_State}
\end{figure}

Based on Eq.~(\ref{Floquetstate}),
we provide an intuitive physical picture to understand the spin dynamics under periodic driving.
As shown in Fig.~\ref{Floquet_State}(a),
two Floquet states $|\uparrow\rangle_n$ and $|\downarrow\rangle_m$ are the upper and lower energy levels, respectively.
The energy difference between them is $E_{\uparrow,n}-E_{\downarrow,m}=\nu_0+(n-m)\nu_{\rm{ac}}$.
Figure~\ref{Floquet_State}(a) shows that a Floquet state can be represented as the superposition state of the particle number representation $|\epsilon, n\rangle$.
The transition between different $|\epsilon, n\rangle$ corresponds to a specific AC-photon number.
For example, $|\downarrow,\bar{N}-1\rangle\rightarrow|\uparrow,\bar{N}-2\rangle$ absorbs one AC photon; $|\downarrow,\bar{N}-1\rangle\rightarrow|\uparrow,\bar{N}-1\rangle$ has no change in the number of AC photons; $|\downarrow,\bar{N}-1\rangle\rightarrow|\uparrow,\bar{N}\rangle$,$|\downarrow,\bar{N}\rangle\rightarrow|\uparrow,\bar{N}+1\rangle$ and $|\downarrow,\bar{N}+1\rangle\rightarrow|\uparrow,\bar{N}+2\rangle\cdots$ emit one AC photon. 

When a measured field oscillates at the frequency of $\nu_0+(n-m)\nu_{\rm{ac}}$,
it induces the transition between $|\uparrow\rangle_n$ and $|\downarrow\rangle_m$ states.
As discussed above, such a Floquet transition is the collective of transitions between different particle number representations $|\epsilon, n\rangle$.
In the following, we focus on the spin dynamics because AC photons are not accessible in our experiment.
As shown in Fig.~\ref{Floquet_State}(b),
the frequency of the resonance signal is $\nu_0+k\nu_{\rm{ac}}$ (here $k=n-m$),
which must be equal to the energy level difference of Floquet states.
It can be seen from Fig.~\ref{Floquet_State}(b) that a spin transition (solid line) is accompanied by a $l$-AC-photon (wavy line).
Because the energy should be conserved as $\nu_0+k\nu_{\rm{ac}}$,
$|l|$-AC-photon assisted transition requires that the transition frequency of the spin should be $\nu_0+(k+l)\nu_{\rm{ac}}$. 
For each spin transition at a specific frequency,
the transition probability comes from the contribution of all the $l$-AC-photon assisted transitions represented by the black parallel lines in Fig.~\ref{Floquet_State}(a).
The corresponding transition amplitude is proportional to 
\begin{equation}
    \sum_{m'-n'=l} {J}_{n-n'} (\frac{\gamma B_{\textrm{ac}}}{2\nu_{\textrm{ac}}}){J}_{m-m'} (\frac{-\gamma B_{\textrm{ac}}}{2\nu_{\textrm{ac}}})={J}_{n-m+l} (\frac{\gamma B_{\textrm{ac}}}{\nu_{\textrm{ac}}})={J}_{k+l} (\frac{\gamma B_{\textrm{ac}}}{\nu_{\textrm{ac}}})={J}_{k+l}(u),
    \label{transition_amp}
\end{equation}
where $u=\frac{\gamma B_{\textrm{ac}}}{\nu_{\textrm{ac}}}$ is the modulation index and $k=n-m$.

As mentioned above, it is only possible to observe the spin rather than photon dynamics in our experiment.
We can find that the Fourier amplitude spectrum of the spin has multiple peaks similar to the $\delta$ function.
As shown in Fig.~\ref{Floquet_State}(c),
the height of the peak corresponds to the transition amplitude, which is proportional to ${J}_{k+l} (\frac{\gamma B_{\textrm{ac}}}{\nu_{\textrm{ac}}})={J}_{k+l}(u)$.
In Sec.~\ref{sec2}, we use the semi-classical Bloch equation method to derive the explicit formula [see Eq.~(\ref{H21_1})].

\section{Derivation of Floquet amplification}
\label{sec2}

This section presents the magnetic resonance of periodically driven (Floquet) spins and the response of Floquet spins to external magnetic fields.
To do this,
we employ an approach based on Bloch equations.
Without a periodic driving,
$^{129}$Xe spins are two-level systems in a bias field.
In this case, the $^{129}$Xe spins are sensitive to the magnetic field oscillating at the $^{129}$Xe Larmor frequency.
However, under an AC field as a periodic driving on $^{129}$Xe,
the response of driven $^{129}$Xe spins is different from that of the system that is not driven~\cite{jiang2021search,su2021search}.
As shown below,
the driven $^{129}$Xe spins can simultaneously amplify the external magnetic fields that oscillate at frequencies of transitions between Floquet states.
The present amplification phenomena of Floquet systems are collectively named ``Floquet amplification''.
We explain the details as follows.

\subsection{Bloch equations}
\label{sec2A}

$^{129}$Xe noble gas is our studied two-level system,
which spatially overlaps with $^{87}$Rb gas in the same vapor cell.
A bias magnetic field $B_{\rm 0}$ is applied along $\hat{z}$.
To periodically drive the spins into a Floquet system,
we introduce an AC field $B_{\rm ac}{\rm cos}(2\pi\nu_{\rm ac}t) \hat{z}$ that is parallel to the bias field.
In addition, there is an oscillating magnetic field $B_{y}\cos(2\pi\nu t)$ to be measured.
Thus, \mbox{$\textbf{B}=B_{y}\cos(2\pi\nu t)\hat{y}+[B_0+B_{\rm{ac}}\cos(2\pi \nu_{\rm{ac}}t)]\hat{z}$} is the total magnetic field experienced by $^{129}$Xe spins.
The spin dynamics of the overlapping $^{129}$Xe and $^{87}$Rb ensembles can be described by the coupled Bloch equations~\cite{kornack2005nuclear},
\begin{eqnarray}
\label{H1}
\frac{\partial \textbf{P}^{e}}{\partial t}&=&\frac{\gamma_{e}}{Q} (\textbf{B}+\lambda {M}^{n}\textbf{P}^{n})\times\textbf{P}^{e} + \frac{P^{e}_{0} \bm{z} -\textbf{P}^{e} }{\{T_{2e},T_{2e},T_{1e}\} Q},\\
\label{H2}
\frac{\partial \textbf{P}^{n}}{\partial t}&=&\gamma_{n}(\textbf{B}+\lambda {M}^{e}\textbf{P}^{e})\times\textbf{P}^{n} +  \frac{P^{n}_{0}\bm{z}-\textbf{P}^{n} }{\{T_{2n},T_{2n},T_{1n}\}},
\end{eqnarray}
where $\gamma_{e} \approx 2\pi\times28~\textrm{Hz/nT}$ and $\gamma_{n} \approx 2\pi\times 0.01178~\textrm{Hz/nT}$ are, respectively, the gyromagnetic ratio of a bare electron and $^{129}$Xe nuclear spin, $\textbf{P}^{e}$ and $\textbf{P}^{n}$ are, respectively, the polarization of $^{87}$Rb electrons and $^{129}$Xe atoms, 
$P_{0}^{e}$ and $P_{0}^{n}$ are the equilibrium polarization of $^{87}$Rb and $^{129}$Xe,
$M^{e}$ and $M^{n}$ are, respectively, the magnetization of $^{87}$Rb atoms and $^{129}$Xe atoms with unity polarization,
$T_{1n},T_{2n} \approx 34$~s are the longitudinal and transverse relaxation times of $^{129}$Xe spins,
and $\lambda M \textbf{P}$ is the effective field induced by $\textbf{P}$ polarization. $T_{1e},T_{2e}$ are the longitudinal and transverse relaxation times of $^{87}$Rb spins.
The relaxation times of $^{87}$Rb spins satisfy $1/T_{2e}=1/T_{1e}+1/T_{2}^{\rm{SE}}$, where $1/T_{2}^{\rm{SE}}$ is the relaxation caused by spin-exchange collisions.
In our experiment, $1/T_{1e}$ is the dominant relaxation term.
Accordingly, $T_{1e} \approx T_{2e}$ can be regarded as the common relaxation time without distinction here.
The factor $Q$ is the slowing-down factor of $^{87}$Rb atoms, which depends on the $^{87}$Rb polarization.
Because $P_{z}^{e}$ is much larger than the transverse components $P_{x}^{e}$ and $P_{y}^{e}$, $Q$ primarily depends on $P_{z}^{e}$.
Moreover, $Q$ can be approximated as a constant because $P_{z}^{e}$ can be approximated as a constant.

In the spatially overlapping $^{129}$Xe and $^{87}$Rb ensembles,
there exists Fermi-contact interaction between $^{129}$Xe and $^{87}$Rb spins.
$^{129}$Xe spins generate an effective field $\lambda M^{n} \textbf{P}^{n}$ on $^{87}$Rb spins~\cite{walker1997spin}.
Such an effective field is 
\begin{equation}
	\textbf{B}_{\rm{eff}}=\lambda M^{n} \textbf{P}^{n}=\frac{8\pi}{3} \kappa_0 M^{n} \textbf{P}^{n},
	\label{beff}
\end{equation}
 where $\kappa_{0}\approx 540$ is the Fermi-contact enhancement factor.
 In order to derive the effective field,
 the core is to determine $\textbf{P}^{n} $.
Similarly, $^{87}$Rb magnetization also generates an effective field $\lambda M^{e} \textbf{P}^{e}$ on $^{129}$Xe spins.
In our experiment, $\lambda M^{e} \textbf{P}^{e}$ is on the order of 1\,nT and $\lambda M^{n} \textbf{P}^{n}$ is at least one order larger than $\lambda M^{e} \textbf{P}^{e}$.
We thus neglect the $\lambda M^{e} \textbf{P}^{e}$ term.
As a result, the coupled Bloch equations, Eqs.\,(\ref{H1},\ref{H2}), can be greatly simplified to
\begin{eqnarray}
\label{H4}
\frac{\partial \textbf{P}^{e}}{\partial t}&=&\frac{\gamma_{e}}{Q} (\textbf{B}+\lambda M^{n} \textbf{P}^{n}) \times\textbf{P}^{e} + \frac{P^{e}_{0} \bm{z} -\textbf{P}^{e} }{T_{e} Q},\\
\label{H5}
\frac{\partial \textbf{P}^{n}}{\partial t}&=&\gamma_{n}\textbf{B}\times\textbf{P}^{n} +  \frac{P^{n}_{0}\bm{z}-\textbf{P}^{n} }{\{T_{2n},T_{2n},T_{1n}\}}.
\end{eqnarray}

Based on Eq.~(\ref{H5}), we note that the evolution of $\textbf{P}^{n}$ is independent on $\textbf{P}^{e}$ and thus can be calculated independently.
We first solve the $\textbf{P}^{n}$ evolution of $^{129}$Xe.
The Larmor frequency of $^{129}$Xe spins can be written as $\nu_{0}=\gamma_{n}B_{0}/(2\pi)$. With rotating-wave approximation, we rewrite the Bloch equation of $^{129}$Xe spins in the rotating frame,
\begin{equation}
\begin{aligned}
\dfrac{\partial \tilde{\textbf{P}}^{n}}{\partial t}=\gamma_{n} \tilde{\textbf{B}} \times \tilde{\textbf{P}}^{n}-\dfrac{\tilde{P}_{x}^{n}\tilde{\bm{x}}+\tilde{P}_{y}^{n}\tilde{\bm{y}}}{T_{2n}}-\dfrac{(P_{z}^{n}-P_{0}^{n})\tilde{\bm{z}}}{T_{1n}},
\label{H6}
\end{aligned}
\end{equation}
where $\tilde{\textbf{B}}=\left[\dfrac{2\pi(\nu_{0}-\nu)}{\gamma_{n}}+B_{\rm{ac}}\cos(2\pi\nu_{\rm{ac}}t)\right]\tilde{\bm{z}}+\dfrac{B_{y}}{2}\tilde{\bm{y}}$ is the equivalent magnetic field in the rotating frame, $\tilde{\bm{x}},\tilde{\bm{y}},\tilde{\bm{z}}$ are the coordinate axes in the rotating frame.
To simplify Eq.~(\ref{H6}),
we introduce matrix notations and obtain
\begin{equation}
	\dfrac{\partial \tilde{\textbf{P}}^n}{\partial t}=(\textbf{$ \Omega $}-\textbf{$ \Gamma $})\tilde{\textbf{P}}^n+\dfrac{P_0^n\tilde{\bm{z}}}{T_{1n}},
	\label{H7}
\end{equation}
where $\Omega$, $\Gamma$, $\tilde{\textbf{P}}^n$ matrices are 
\begin{equation}
	\Omega=
	\left[
	\begin{matrix}
		0 & -\Omega_z & \Omega_y \\
		\Omega_z & 0 & -\Omega_x \\
		-\Omega_y & \Omega_x & 0
	\end{matrix}
	\right],
	\left\{
	\begin{array}{l}
		\Omega_x =0 \\
		\Omega_y =\dfrac{\gamma_n B_y}{2} \\
		\Omega_z =2\pi(\nu_0-\nu)+\gamma_n B_{ac}\cos(2\pi\nu_{\rm{ac}}t)
	\end{array},
	\right.
	\label{H8}
\end{equation}

\begin{equation}
	\begin{matrix}
		\Gamma=
	\left[
	\begin{matrix}
		1/T_{2n} & 0 & 0 \\
		0 & 1/T_{2n} & 0 \\
		0 & 0 & 1/T_{1n}
	\end{matrix}
	\right],
	\end{matrix} 
	\label{H10}
\end{equation}

\begin{equation}
	\tilde{\textbf{P}}^n=\left[
	\begin{matrix}
		\tilde{P}^n_x \\
		\tilde{P}^n_y \\
		\tilde{P}^n_z
	\end{matrix}
	\right],
	\dfrac{P_0^n\tilde{\bm{z}}}{T_{1n}}=
	\left[
	\begin{matrix}
		0  \\
		0 \\
		P_0^n/T_{1n}
	\end{matrix}
	\right].
	\label{H9}
\end{equation}

A way to solve Eq.~(\ref{H7}) is to combine $\tilde{P}^n_x$ and $\tilde{P}^n_y$ into $\tilde{P}^n_{\pm}=\tilde{P}^n_x \pm i \tilde{P}^n_y$, and solve for the evolution of $\tilde{P}^n_{\pm}$.
To do this,
we define a matrix $\Lambda$
\begin{equation}
	\Lambda=
	\left[
	\begin{matrix}
		1 & i & 0 \\
		1 & -i & 0 \\
		0 & 0 & 1
	\end{matrix}
	\right].
	\label{H11}
\end{equation}
Using the $\Lambda$ matrix, Eq.~(\ref{H7}) becomes
\begin{equation}
	\dfrac{\partial}{\partial t}\Lambda\tilde{\textbf{P}}^n=\Lambda(\Omega-\Gamma)\Lambda^{-1}\Lambda\tilde{\textbf{P}}^n+\Lambda\dfrac{P_0^n\tilde{\bm{z}}}{T_{1n}},
	\label{H12}
\end{equation}	
where some terms can be further simplified to
\begin{equation}
	\Lambda\tilde{\textbf{P}}^n=\left[
	\begin{matrix}
		\tilde{P}^n_+ \\
		\tilde{P}^n_- \\
		\tilde{P}^n_z
	\end{matrix},
	\right],
	\Lambda\dfrac{P_0^n\tilde{\bm{z}}}{T_{1n}}=
	\left[
	\begin{matrix}
		0  \\
		0 \\
		P_0^n/T_{1n}
	\end{matrix}
	\right],
	\label{H14}
\end{equation}
\begin{equation}
	\Lambda(\Omega-\Gamma)\Lambda^{-1}=
	\left[
	\begin{matrix}
		-1/T_{2n}+i\Omega_z & 0 & -i\Omega_+ \\
		0 & -1/T_{2n}-i\Omega_z & i\Omega_- \\
		-i\Omega_-/2 & i\Omega_+/2 & -1/T_{1n}
	\end{matrix}
	\right],
	\label{H13}
\end{equation}

\begin{equation}
	\left\{
	\begin{matrix}
		\Omega_+=& \Omega_x+i\Omega_y\\
		\Omega_-=& \Omega_x-i\Omega_y\\
		\tilde{P}^n_+=&\tilde{P}^n_x+i\tilde{P}^n_y \\
		\tilde{P}^n_-=&\tilde{P}^n_x-i\tilde{P}^n_y 
	\end{matrix}.
	\right.
	\label{H15}
\end{equation}
After the above matrix transformation,
we obtain three differential equations from Eq.~(\ref{H13})
\begin{subequations}
	\begin{equation}
	    \dfrac{\partial \tilde{P}^n_+}{\partial t}= \left(-\dfrac{1}{T_{2n}}+i\Omega_z\right)\tilde{P}^n_+-i\Omega_+\tilde{P}^n_z,
	    \label{H16a}
	\end{equation}
	\begin{equation}
	    \dfrac{\partial \tilde{P}^n_-}{\partial t}= \left(-\dfrac{1}{T_{2n}}-i\Omega_z\right)\tilde{P}^n_-+i\Omega_-\tilde{P}^n_z,
	    \label{H16b}
	\end{equation}
	\begin{equation}
	    \dfrac{\partial \tilde{P}^n_z}{\partial t}=-\dfrac{i}{2}\Omega_-\tilde{P}^n_+ +\dfrac{i}{2}\Omega_+\tilde{P}^n_--\dfrac{\tilde{P}^n_z}{T_{1n}}+\dfrac{P^n_0}{T_{1n}}.
	    \label{H16c}
	\end{equation}
\end{subequations}

To solve Eq.~(\ref{H16c}),
we note that in the experiment the measured-field amplitude $B_y$ is small and satisfies the condition $|\Omega_+|=|\Omega_-|=\dfrac{\gamma_n B_y}{2}\ll\dfrac{1}{T_{1n}}$.
Such two terms $-\dfrac{i}{2}\Omega_-\tilde{P}^n_+$ and $\dfrac{i}{2}\Omega_+\tilde{P}^n_-$ can be neglected in Eq.~(\ref{H16c}).
As a result,
$\tilde{P}_z^n$ can be approximated as $\tilde{P}_z^n \approx P_0^n$.
Using this approximation,
the Eq.~(\ref{H16a}) can be written as
\begin{equation}
\dfrac{\partial \tilde{P}^n_+}{\partial t} \approx \left(-\dfrac{1}{T_{2n}}+i\Omega_z\right)\tilde{P}^n_+-i\Omega_+P^n_0 .
\label{H17}
\end{equation}
This differential equation has an analytical solution
\begin{equation}
\tilde{P}^n_+(t)=\overbrace{C\exp\left[\int_{0}^{t}\left(-\dfrac{1}{T_{2n}}+i\Omega_z\right)dt'\right]}^\textrm{Transient response}-\overbrace{\int_{0}^{t}i\Omega_+P^n_0\exp\left[\int_{t'}^{t}\left(-\dfrac{1}{T_{2n}}+i\Omega_z\right)dt''\right]dt'}^\textrm{Steady-state response},
\label{H18}
\end{equation}
where
$\tilde{P}^n_+(t)$ contains two different responses, i.e., transient and steady-state response.
The first term in Eq.~\ref{H18} is transient response,
which is oscillating free decay with the time scale of coherence time $T_{2n}$.
Because our experiment focuses on the steady-state solution, we neglect the transient response.
In the following, we describe the second term (steady-state response).

The steady-state response in Eq.~(\ref{H18}) can be further calculated as
\begin{equation}
\begin{aligned}
\tilde{P}^n_+(t) &= -i\Omega_+P^n_0\int_{0}^{t}\exp\left[\int_{t'}^{t}\left(-\dfrac{1}{T_{2n}}+i\Omega_z\right)dt''\right]dt' = \dfrac{\gamma_n P^n_0 B_y}{2}\int_{0}^{t}\exp\left\{\int_{t'}^{t}-\dfrac{1}{T_{2n}}+i\left[2\pi(\nu_0-\nu)+\gamma_n B_{\rm{ac}}\cos(2\pi\nu_{\rm{ac}}t'')\right]dt''\right\}dt' \\
&=\dfrac{\gamma_n P^n_0 B_y}{2}\int_{0}^{t}\exp\left\{\left[-\dfrac{1}{T_{2n}}+i2\pi(\nu_0-\nu)\right]\left(t-t'\right)+
\overbrace{\dfrac{i\gamma_n B_{\rm{ac}}}{2\pi\nu_{\rm{ac}}}\left[\sin(2\pi\nu_{\rm{ac}}t)-\sin(2\pi\nu_{\rm{ac}}t')\right]}^{\rm{Floquet~term}}\right\}dt'.
\label{H19}
\end{aligned}
\end{equation}
In contrast to the case without periodic driving, $\tilde{P}^n_+(t)$ contains a new term called Floquet term.
To further obtain the solution of $\tilde{P}^n_+(t)$, we utilize the Jacobi$-$Anger identity
\begin{equation}
\exp[iu\, {\sin}(\omega t)]=\sum_{k=-\infty}^{+\infty}J_k(u)\exp(ik\omega t),
\label{H20}
\end{equation}
where $u=\dfrac{\gamma_n B_{\rm{ac}}}{2\pi \nu_{\rm{ac}}}$, $\omega=2\pi\nu_{\rm{ac}}$,
and $J_k$ is the Bessel function of the first kind.
After straightforward calculations,
$\tilde{P}^n_+(t)$ becomes
\begin{equation}
\begin{aligned}
\tilde{P}^n_+(t) &= \dfrac{\gamma_n P^n_0 B_y}{2}\exp\left\{\left[-\dfrac{1}{T_{2n}}+i2\pi(\nu_0-\nu)\right]t\right\}
\sum_{k'=-\infty}^{+\infty}\sum_{k=-\infty}^{+\infty}J_{k'}(u)J_{k}(u)\\
&\times \exp\left(i2\pi k' v_{\rm{ac}}t\right)\int_{0}^{t}\exp\left\{-\left[-\dfrac{1}{T_{2n}}+i2\pi(\nu_0-\nu)\right]t'-i2\pi k v_{\rm{ac}}t'\right\}\\
&=\dfrac{\gamma_n P^n_0 B_y}{2}\sum_{k'=-\infty}^{+\infty}\sum_{k=-\infty}^{+\infty}J_{k'}(u)J_{k}(u)\dfrac{\exp\left[i2\pi \left(k'-k\right)v_{\rm{ac}}t\right]}{\dfrac{1}{T_{2n}}-i2\pi(\nu_0-\nu)-i2\pi k v_{\rm{ac}}}\\
&=\dfrac{\gamma_n P^n_0 T_{2n} B_y}{2}\sum_{l=-\infty}^{+\infty}\exp\left(i2\pi l v_{\rm{ac}}t\right)\sum_{k=-\infty}^{+\infty}J_{k+l}(u)J_{k}(u)\dfrac{1+i2\pi\left(\nu_0-\nu+ kv_{\rm{ac}}\right) T_{2n}}{1+\left[2\pi\left(\nu_0-\nu+k v_{\rm{ac}}\right) T_{2n}\right]^2}.
\label{H21}
\end{aligned}
\end{equation}
We transform $\tilde{P}^n_+(t)$ from the rotating coordinate back to the laboratory coordinate $P^n_+(t)=P^n_x+iP^n_y$,

\begin{equation}
\begin{aligned}
{P}^n_+(t)=\dfrac{\gamma_n P^n_0 T_{2n} B_y}{2}\sum_{l=-\infty}^{+\infty}\exp [i2\pi (l v_{\rm{ac}} + \nu)t] \sum_{k=-\infty}^{+\infty}J_{k+l}(u)J_{k}(u)\dfrac{1+i2\pi\left(\nu_0-\nu+ kv_{\rm{ac}}\right) T_{2n}}{1+\left[2\pi\left(\nu_0-\nu+k v_{\rm{ac}}\right) T_{2n}\right]^2}.
\label{H21_1}
\end{aligned}
\end{equation}
This result is essential in Floquet amplification.
We summarize the main difference between the driven and undriven cases in Table~\ref{table1}.

\begin{table}[htb]
\begin{center}
\label{difference10}
    \renewcommand\arraystretch{1.5}
    \caption{Comparison between the driven case and the undriven case}
    \label{table1}
    \setlength{\tabcolsep}{11mm}{
    \begin{tabular}{ccc}
    \toprule
     ~&Driven case&Undriven case  \\
     \hline
     modulation index $u$ & $>0$ & 0 \\
     \hline
     resonance frequency &$\nu\approx \nu_0+k\nu_{\rm{ac}}$&$\nu \approx \nu_{0}$ \\
     \hline
     $l$-AC-photon transition amplitude&$ \propto J_{k+l}(u)J_{k}(u) $& No AC-photon transition \\
     \hline
\end{tabular}
    }
\end{center}
\end{table}

\subsection{Floquet amplification}
\label{sec2B}

We now derive the Floquet amplification factors.
Based on $P^n_+(t)$ in Eq.~(\ref{H21_1}),
we can obtain the steady-state solution of the spin polarization along $\hat{x}$ and $\hat{y}$,
\begin{eqnarray}
    \label{H22}
    P^n_x(t)=B_y \sum_{l=-\infty}^{+\infty}\sum_{k=-\infty}^{+\infty}A_{k,l}(u,\nu)\cos[2\pi(\nu+l\nu_{\rm{ac}})t]-B_{k,l}(u,\nu)\sin[2\pi(\nu+l\nu_{\rm{ac}})t],\\
    \label{H23}
    P^n_y(t)=B_y \sum_{l=-\infty}^{+\infty}\sum_{k=-\infty}^{+\infty}B_{k,l}(u,\nu)\cos[2\pi(\nu+l\nu_{\rm{ac}})t]+A_{k,l}(u,\nu)\sin[2\pi(\nu+l\nu_{\rm{ac}})t],
\end{eqnarray}
where the coefficients $A_{k,l}(u,\nu)$ and $B_{k,l}(u,\nu)$ are
\begin{eqnarray}
    \label{H24}
    A_{k,l}(u,\nu)=\dfrac{\gamma_n P^n_0 T_{2n}  J_{k+l}(u)J_{k}(u)}{2}\dfrac{1}{1+\left[2\pi\left(\nu_0-\nu+k v_{\rm{ac}}\right) T_{2n}\right]^2},\\
    \label{H25}
    B_{k,l}(u,\nu)=\dfrac{\gamma_n P^n_0 T_{2n}  J_{k+l}(u)J_{k}(u)}{2}\dfrac{2\pi\left(\nu_0-\nu+ kv_{\rm{ac}}\right) T_{2n}}{1+\left[2\pi\left(\nu_0-\nu+k v_{\rm{ac}}\right) T_{2n}\right]^2}.
\end{eqnarray}
The physical meaning of these equations is that the transition amplitudes are proportional to the products of the amplitudes of the initial and final Floquet states and a resonant Lorentz factor.

In Sec.\,\ref{sec2A}, we have discussed that the spin polarization $P^n_x(t)$ and $P^n_y(t)$ of $^{129}$Xe can generate an effective magnetic field on $^{87}$Rb atoms; see Eq.~(\ref{beff}).
Based on Eqs.~(\ref{beff}, \ref{H22}-\ref{H25}),
we can obtain the ratio between the effective field and the measured field, i.e., Floquet amplification factors.
Here, because $^{87}$Rb magnetometer is primarily sensitive to the magnetic field along $\hat{y}$,
we only need consider the $B_{\rm{eff,y}}=\frac{8\pi}{3}\kappa_0 M^n P^n_y(t)$.
The amplification factor is
\begin{equation}
\eta(u,\nu) \equiv \dfrac{B_{\rm{eff},y}}{B_y}=\sqrt{\left(\sum_{k=-\infty}^{+\infty}A_{k,l}(u,\nu)\right)^2+\left(\sum_{k=-\infty}^{+\infty}B_{k,l}(u,\nu)\right)^2}.
\label{H26}
\end{equation}

\begin{figure}[t]  
	\makeatletter
\centering
	\def\@captype{figure}
	\makeatother
	\includegraphics[scale=1.1]{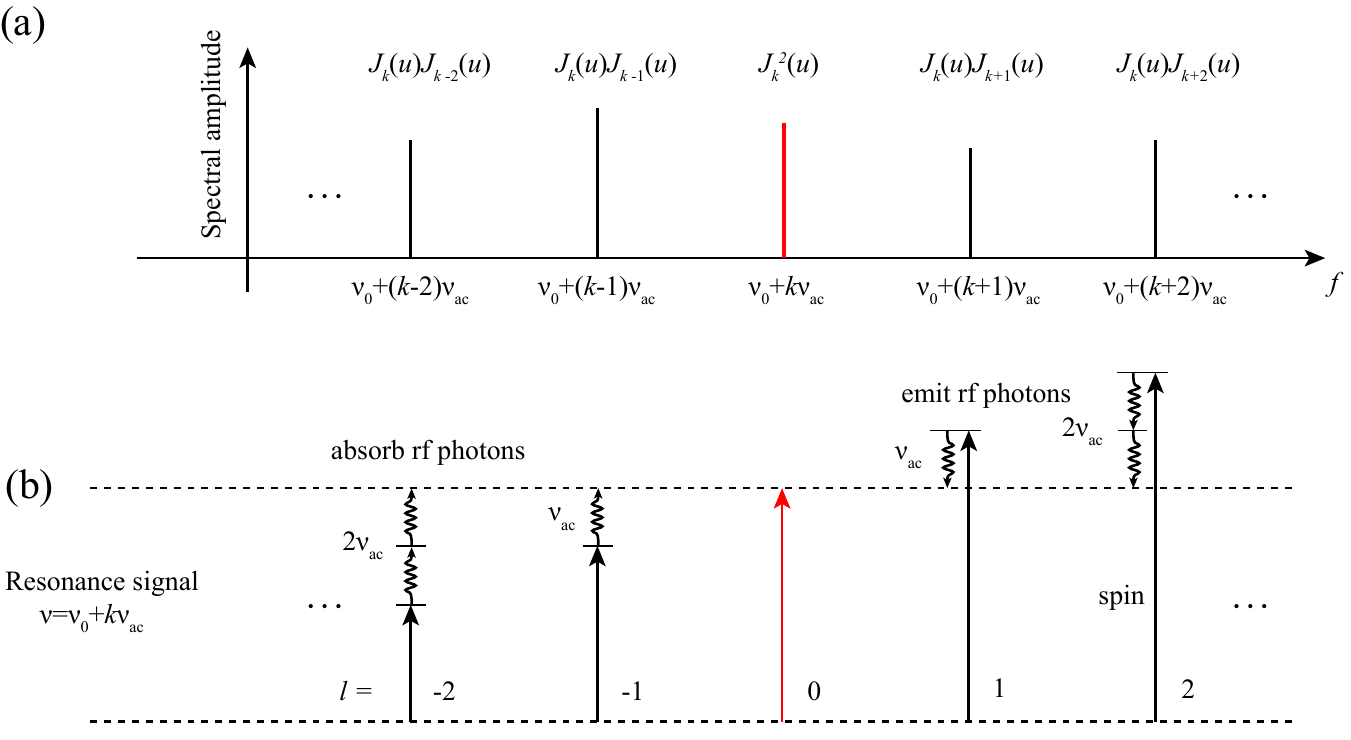}
	\caption{Floquet state and multiple AC photon transitions.}
	\label{Floquet_State}
\end{figure}

When the oscillation frequency of the measured field matches with one of Floquet transitions,
the amplification factor becomes significant.
We consider the case in which the frequency of the measured field satisfies $\nu_0-\nu+k v_{\rm{ac}}\approx 0$.
The amplification factor $\eta_{k,l}(u)$ becomes
\begin{equation}
\eta_{k,l}(u)=\dfrac{4\pi}{3}\kappa_0 M^n P^n_0\gamma_n T_{2n} J_{k+l}(u) J_{k}(u).
\label{H27}
\end{equation}
Here $k$ and $l$ are used to describe the frequencies of the measured and amplified fields.
Specifically, $k$ denotes that the measured field frequency is at $\nu=\nu_0+k \nu_{\rm{ac}}$,
and $l$ denotes that the output signal frequency is at $\nu_0+(k+l) \nu_{\rm{ac}}$.
There is a particular but common example, when $l=0$, the measured field and amplified signal are both at $\nu=\nu_0+k \nu_{\rm{ac}}$.
In this case, the amplification factor is given by
\begin{equation}
\eta_{k,0}(u)=\dfrac{4}{3}\kappa_0 M^n P^n_0\gamma_n T_{2n}J^2_{k}(u).
\label{H29}
\end{equation}
For the case of $l \neq 0$,
the applied and detected frequencies are different.
This kind of amplification originates from the fact that, although the test frequency matches only one Floquet transitions, there simultaneously exist amplification signals at other Floquet transitions.
For example shown in Fig.~2(b) of the main text,
there is a peak at the frequency of the test field set at 11.539 Hz (marked with a star),
and otherwise there exist other sidebands even with larger signal amplitude.

\subsection{Multiple resonance}
\label{sec2C}

\begin{figure}[htb]  
	\makeatletter
\centering
	\def\@captype{figure}
	\makeatother
	\includegraphics[scale=0.55]{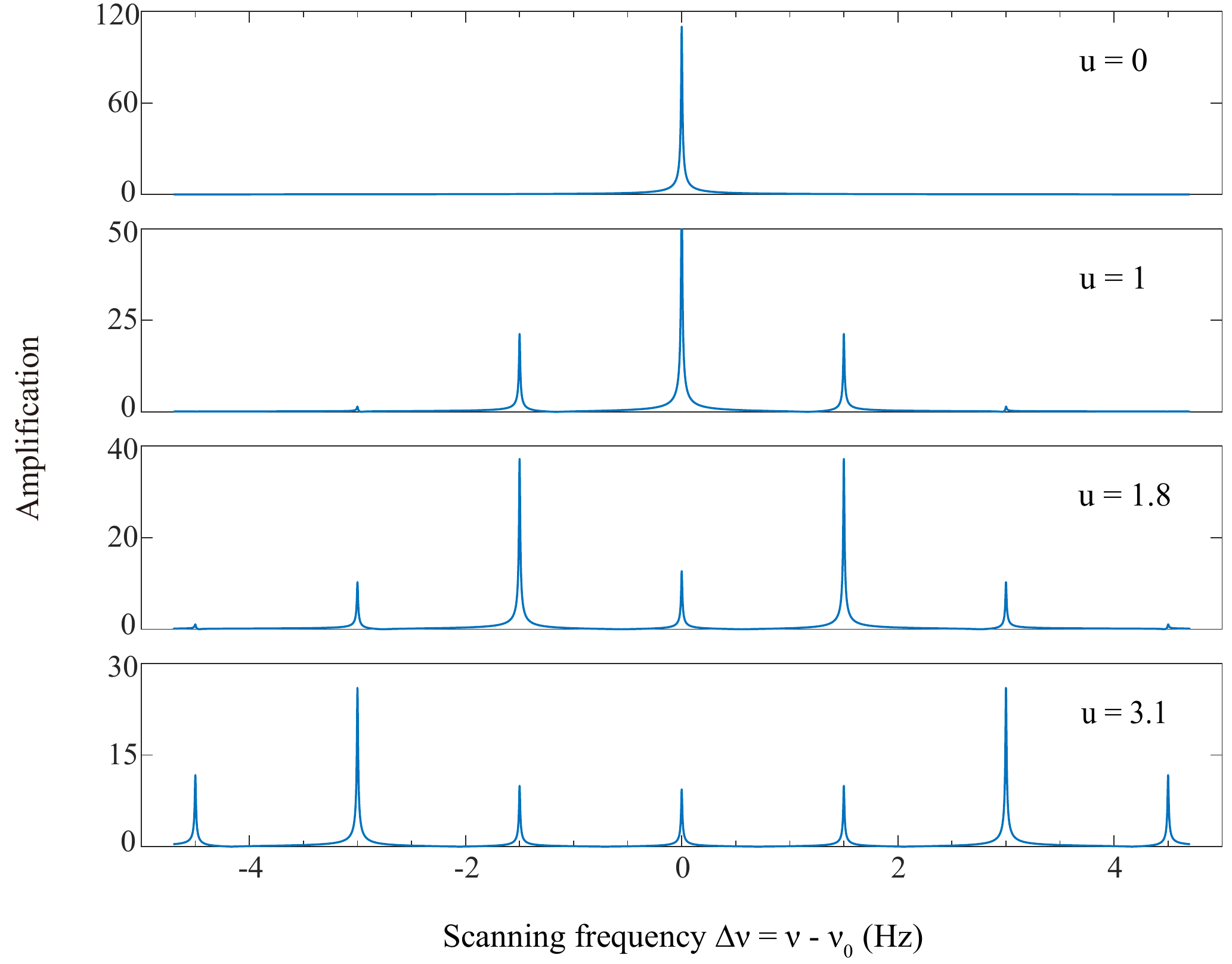}
	\caption{Plot of amplification versus modulation index $ u = \gamma B_{\rm ac}/ \nu_{\rm ac} $. The bias field is set as $B_{0}\approx853$ nT and the frequency of periodic driving field is $\nu_{ \rm{ac}}\approx 1.500$ Hz. We change the amplitude of $B_{\rm ac}$ to realize different value of $u$. As shown above, the number of Floquet amplification peaks depends on the specific value of $u$. }
	\label{bandwith}
\end{figure}

We now calculate the frequency bandwidth of Floquet amplification.
Based on Eqs.~(\ref{H27}) and (\ref{H29}),
the measured frequency $\nu$ should satisfy $\nu \approx \nu_0+k\nu_{\rm{ac}}$,
where $k$ is an integer,
otherwise the amplification factor is small.
The profile of the amplification is
\begin{equation}
\eta(u,\nu)=\sum_k J_{k}^2(u) \dfrac{\dfrac{1}{2}\lambda M^n P^n_0\gamma_n T_{2n}}{\sqrt{1+\left[2\pi\left(\nu_0-\nu+k v_{\rm{ac}}\right) T_{2n}\right]^2}},
\label{H30}
\end{equation}
which is a sum of resonance with the full-width at half-maximum (FWHM) for each amplification regime being $\sqrt{3}/\pi T_{2n}$.
The FWHM is measured to be about 17~mHz.
We plot the amplification profile under different modulation index $u$ in Fig.~\ref{bandwith}.
For example, when the modulation index is chosen as $u=3.1$,
the amplification profile contains seven peaks,
allowing to detect the signal at seven different regimes.
In contrast, when there is no periodic driving corresponding to $u=0$,
there is only one amplification peak,
limiting the detection regimes.

\subsection{Power amplification}
\label{Subsec:Power_Ampl}

\begin{figure}[htb]  
	\makeatletter
\centering
	\def\@captype{figure}
	\makeatother
	\includegraphics[scale=0.78]{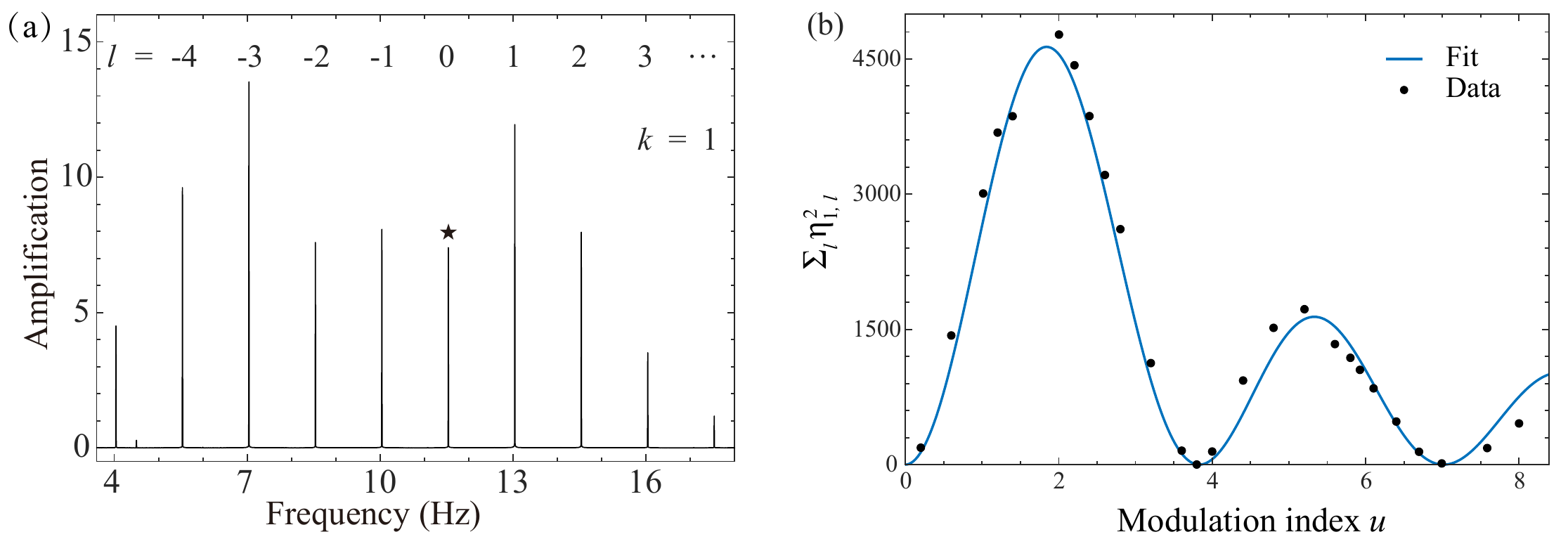}
	\caption{(a) Spectrum of the amplification signal induced by a test field. The test signal is set at the frequency of 1st-order sideband (see star) and there simultaneously exists other sidebands signals. The amplification satisfies $\eta_{k,l}$. (b) Sum of the squares of amplification factors $\sum_l\eta^2_{1,l}$ and fitting.}
	\label{sec4figure}
\end{figure}

When a signal is applied to the Floquet amplifier on a Floquet resonance, the device outputs signals at a set of output frequencies, see Fig.~\ref{sec4figure}(a). 
There exists an important sum rule for the power gain over all the output resonances. Indeed,
since
\begin{equation}
\sum_{l=-\infty}^{+\infty}J_l^2(u)=1,
\end{equation}
we have
\begin{equation}
\sum_{l=-\infty}^{+\infty}\eta_{k,l}^2(u)=\sum_{l=-\infty}^{+\infty}\left(\dfrac{1}{2}\lambda M^n P^n_0\gamma_n T_{2n}\left|J_{k+l}(u)J_{k}(u)\right|\right)^2=\left(\dfrac{1}{2}\lambda M^n P^n_0\gamma_n T_{2n}\right)^2J_k^2(u)\sum_{l=-\infty}^{+\infty}J_{k+l}^2=\eta_{0,0}(0)\eta_{k,0}(u).
\end{equation}
These predictions are verified by our measurements as shown in Fig.\,\ref{sec4figure}(b).

\section{Fano resonance}
\label{sec3}

We find that the line shape of Floquet amplification as a function of the test frequency is not symmetric [see Fig.\,2(a) in the main text].
The asymmetric profile is significantly distorted compared to the symmetric shape of Eq.\,\eqref{H30}.
As discussed below,
the asymmetric line shape is caused by the well-known Fano resonance.
Specifically,
$\rm ^{87}Rb$ and $\rm ^{129}Xe$ spins both experience the measured oscillating field and both generate signatures at the frequency of the measured field.
The two responses interfere causing the observed asymmetric line shape.
We explain the details as follows.

\begin{figure}[htb]  
	\makeatletter
\centering
	\def\@captype{figure}
	\makeatother
	\includegraphics[scale=1.0]{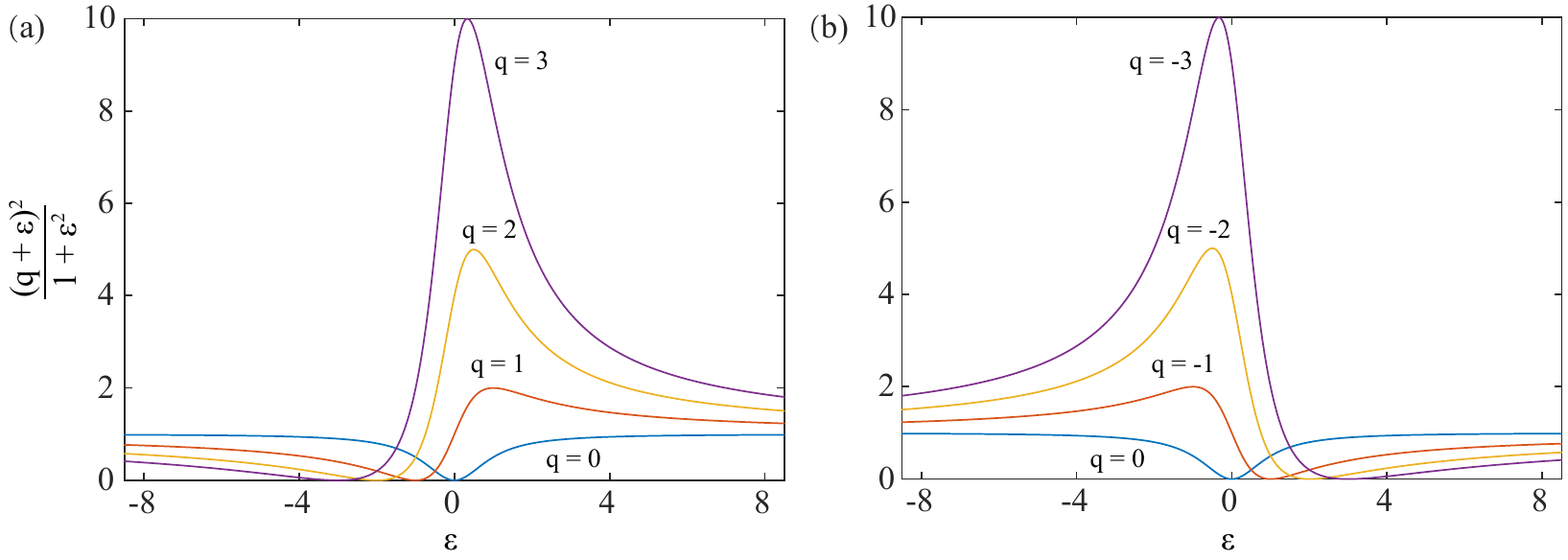}
	\caption{Fano line shape for different values of the Fano parameter $q$.}
	\label{Fano}
\end{figure}


We first introduce the physics of Fano resonance,
which is a ubiquitous scattering wave phenomenon commonly found in a broad range of science and engineering including atomic and solid-state physics, nonlinear optics, photonics, and electronic circuits, etc.
In 1961, Ugo Fano published a paper in \emph{Physical Review}~\cite{fano1961effects}, but the original results for an important limiting case appeared back in 1935.
In general, the Fano resonance occurs when a process involving
a discrete spectrum interferes with a process involving a continuum.
The line shape of the resonance $\sigma(E)$ is described by the Fano formula
\begin{equation}
    \sigma(E)=\overbrace{\sigma_a(E)\dfrac{(q+\epsilon)^2}{1+\epsilon^2}}^{\rm{asymmetric}}+ \overbrace{\sigma_b(E)}^{\rm{symmetric}},
    \label{35}
\end{equation}
where $\sigma_a(E)$ and $\sigma_b(E)$ respectively represent the cross section that does (does not) involve the discrete system,
\begin{equation}
\epsilon=\frac{E-E_r}{\Gamma/2}, 
\end{equation}
$E_r$
is resonance energy, $\Gamma$ is the the natural line width of the discrete system.
$\sigma(E)$ thus generally contains an asymmetric part and a symmetric part.

The Fano parameter, $q$, is important to determine the asymmetry of line shape, according to Eq.\,\eqref{35}, as shown in Fig.\,\ref{Fano}.
The line shape is symmetric for the case of $q=0$.
In contrast, when the Fano parameter $q$ is positive or negative, the shapes show obvious asymmetry.
Moreover, the comparison between the cases of $q>0$ and $q<0$ shows that the line shape also depends on the sign of the Fano parameter.

\begin{figure}[htb]  
	\makeatletter
\centering
	\def\@captype{figure}
	\makeatother
	\includegraphics[scale=0.9]{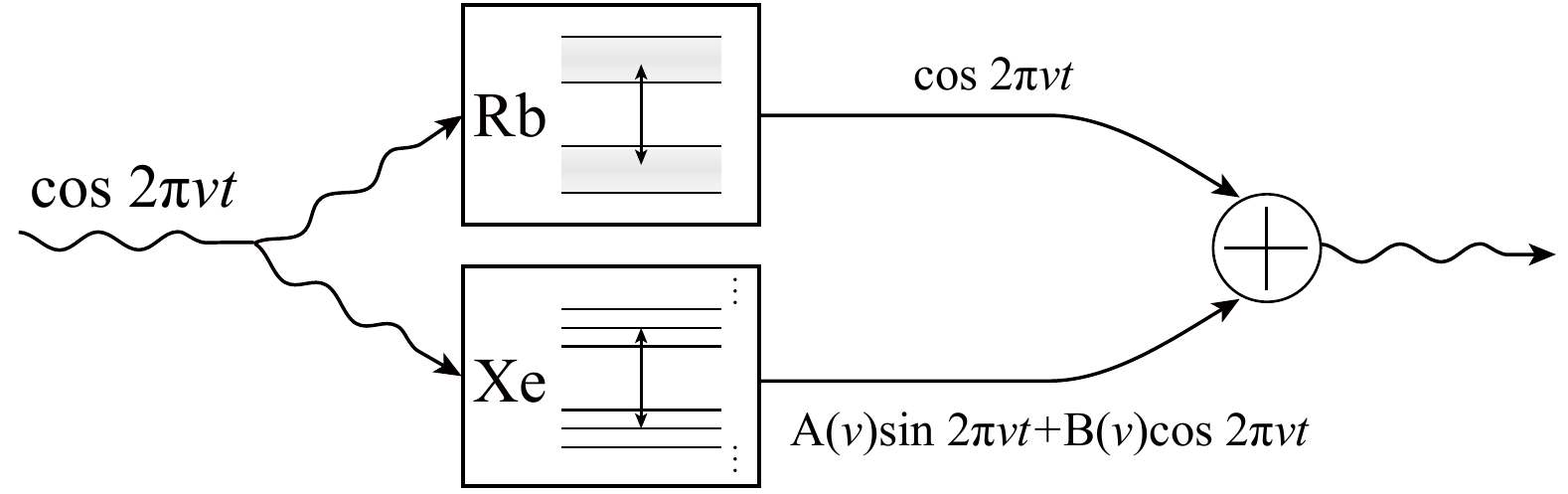}
	\caption{Fano resonance between $^{87}\rm{Rb}$ and $^{129}\rm{Xe}$ systems.}
	\label{Fanointerference}
\end{figure}

To connect our experiment with the Fano resonance,
we first consider what the discrete system and the continuum system are and what the Fano parameter is in our case.
In the $^{129}$Xe-$^{87}$Rb vapor cell,
$\rm ^{129}Xe$ spins have a sharp resonance line due to the long coherence time ($T_{2n}\approx 34$\,s) and thus can be considered as the discrete system.
In contrast to $\rm ^{129}Xe$ spins, $\rm ^{87}Rb$ spins have a broader resonance line due to the shorter coherence time ($T_{2e}\sim$1\,ms) and can be approximated as a continuum system.
As shown in Fig.\,\ref{Fanointerference},
$\rm ^{87}Rb$ and $\rm ^{129}Xe$ spins both experience the measured oscillating field and both generate signatures at the frequency of the measured field.
The phase of $\rm ^{129}Xe$ induced signal changes rapidly near resonance,
while the phase of $\rm ^{87}Rb$ direct signal varies more slowly than that of $\rm ^{129}Xe$.
Two signatures from $\rm ^{87}Rb$ and $\rm ^{129}Xe$ interfere with each other and give rise to the asymmetric Fano shape.

We now derive the Fano formula of the line shape of our signal.
Considering the measured field $B_y\cos(2\pi\nu t)\hat{y}$ and the response $P^n_y(t)$ [see Eq.~(\ref{H23})] of $^{129}\rm{Xe}$,
the amplitude of the output signal relative to the amplitude of the measured field is 
\begin{equation}
\eta(u,\nu)=\sqrt{\left(\sum_{k=-\infty}^{+\infty}A_{k,0}(u,\nu)\right)^2+\left(\sum_{k=-\infty}^{+\infty}B_{k,0}(u,\nu)+1\right)^2}.
\label{H32}
\end{equation}
In the following, we focus on the square of $\eta$, i.e., $\eta ^2(u,\nu)$
and show that the line shape of $\eta ^2(u,\nu)$ as a function of $\nu$ can be described by the Fano function.
For the sake of simplicity, we calculate the $k$-th sideband and ignore the other terms in the summation term of the Eq.\,\eqref{H32}
\begin{equation}
\begin{aligned}
 \eta^2(u,\nu)=\left \{ \dfrac{\eta_{k,0}}{1+\left[2\pi\left(\nu_0-\nu\right) T_{2n}\right]^2}\right \}^2+\left \{\dfrac{2\pi\left(\nu_0-\nu\right) T_{2n}\eta_{k,0}}{1+\left[2\pi\left(\nu_0-\nu\right) T_{2n}\right]^2}+1\right \}^2
 =\overbrace{\dfrac{\left[-\eta_{k,0}+2\pi\left(\nu-\nu_0\right) T_{2n}\right]^2}{1+\left[2\pi\left(\nu-\nu_0\right) T_{2n}\right]^2}}^{\rm{asymmetric}}+ \overbrace{\dfrac{1}{1+\left[2\pi\left(\nu-\nu_0\right) T_{2n}\right]^2}}^{\rm{symmetric}},
    \label{H34}
\end{aligned}
\end{equation}
where $\eta_{k,0}$ is the amplification factor with the modulation index $u$.
Comparing with the standard Fano function in Eq.\,\eqref{35},
we obtain the Fano parameter $q$ for our case
\begin{equation}
q=-\eta_{k,0},
\end{equation}
and $\Gamma=1/\pi T_{2n}$.
As a result, we verify that the asymmetric line shape is indeed caused by the Fano resonance between $^{129}$Xe and $^{87}$Rb.
The profile shown in the main text [Fig.~2a] is the amplitude of the amplification,
and further the square of the amplitude function is the Fano profile.

\begin{figure}[t]  
	\makeatletter
\centering
	\def\@captype{figure}
	\makeatother
	\includegraphics[scale=1.4]{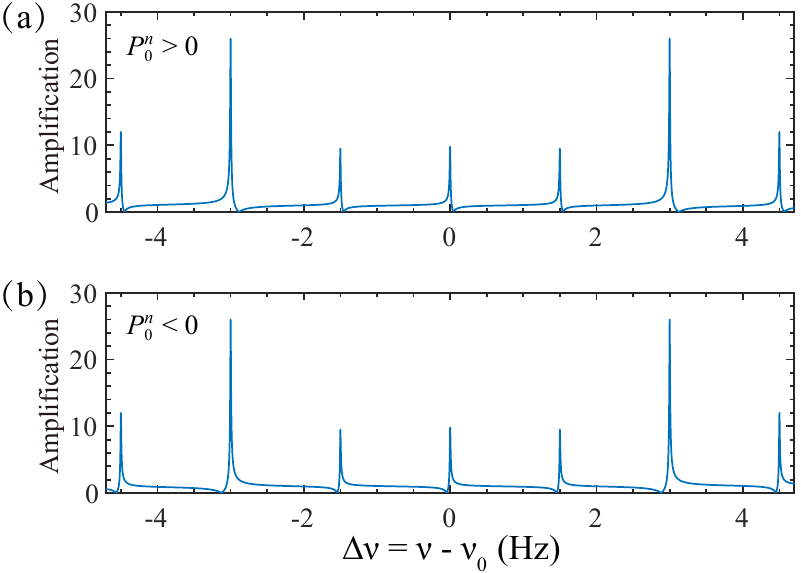}
	\caption{The simulation of the amplification factor profile under different polarization directions. (a) $P^n_0>0$; (b) $P^n_0<0$.}
	\label{sec3figure1}
\end{figure}

The sign of the Fano parameter $q$ depends on the sign of $\eta_{k,0}$, which is important to determine the asymmetry of the line shape.
On the other hand,
based on Eq.~(\ref{H27}),
the sign of $\eta_{k,0}$ can be positive or negative,
depending on the sign of spin polarization, gyromagnetic ratio, etc.
As shown in Fig.\,\ref{sec3figure1}, we simulated the cases where $P^n_0$ is positive and negative, respectively, with $|\eta_{0,0}(0)|=110$ and $T_{2n}=34$ s.
The profile of the experimental data is consistent with the Fig.~\ref{sec3figure1}(b) diagram with $P^n_0<0$.
When we reverse the direction of the pump light or equivalently reverse the direction of $B_0$,
the profile can be observed the same as Fig. \ref{sec3figure1}(a), whose amplification factor is less than 1 on the right side of the peak.

\begin{figure}[htb]  
	\makeatletter
\centering
	\def\@captype{figure}
	\makeatother
	\includegraphics[scale=1.1]{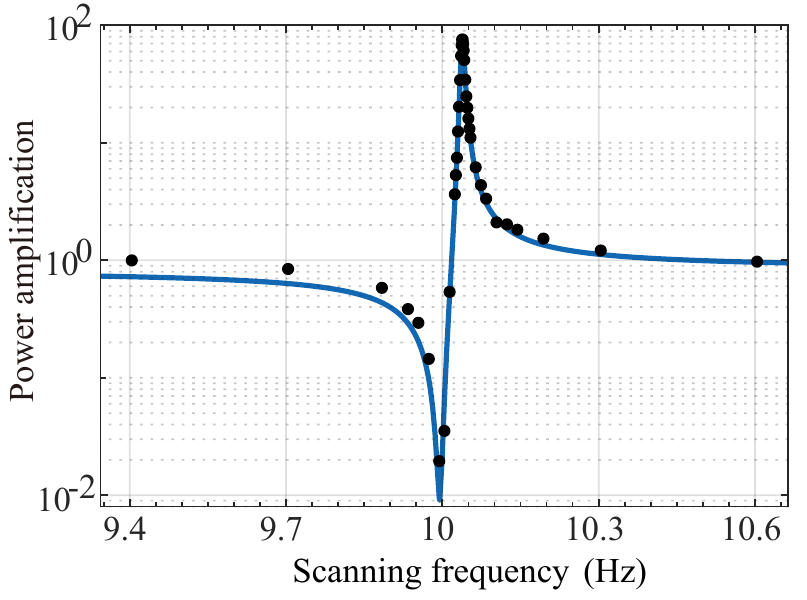}
	\caption{Power amplification of the central sideband. The profile of the power amplification conforms to the Fano line shape. Note the logarithmic vertical scale.}
	\label{sec3figure2-3}
\end{figure}

Based on the above discussions,
we fit the experimental profile (see Fig.\,2a in the main text) with the Fano function.
In experiment, we observe seven lines in the amplification profile.
Based on Eqs.\,\eqref{H24} and \eqref{H25},
we take the explicit forms of $A_{k,0}(u,\nu)$ and $B_{k,0}(u,\nu)$ into the $\eta(u,\nu)$ and 
obtain the fit function
\begin{equation}
		f(\nu)\approx \sum\limits_{k=-3}^{3} \left[{\dfrac{\eta_{k,0}}{1+\left(2\pi(\nu_k-\nu)T_{2n}\right)^2}}\right]^2+\left[{\dfrac{\eta_{k,0} 2\pi(\nu_k-\nu)T_{2n}}{1+\left(2\pi(\nu_k-\nu)T_{2n}\right)^2}}+1\right]^2,
		\label{H33}
\end{equation}
where $\eta_{k,0},\nu_k$ ($k$ is from -3 to 3),
$T_{2n}$, $c$ are fit parameters.
The fitting results are shown in Table~\ref{table2}, and the corresponding fit is shown in Fig.\,\ref{sec3figure2-3}.
The amplification factors are negative and thus the Fano parameter $q_k=-\eta_{k,0}$ is positive.
We have demonstrated the line shape for the case of $q_k>0$ in Fig.\,\ref{Fano}.
The experimental data are in agreement with theoretical analysis.

\begin{table}[htb]
\begin{center}
    \renewcommand\arraystretch{1.5}
    \caption{The fitting result of the data of the amplification factor}
    \label{table2}
    \setlength{\tabcolsep}{4mm}{
    \begin{tabular}{cccccccc}
    \toprule
     $\eta_{-3,0}$&$\eta_{-2,0}$&$\eta_{-1,0}$&$\eta_{0,0}$&$\eta_{1,0}$&$\eta_{2,0}$&$\eta_{3,0}$&$T_{2n}$  \\
     \hline
     -12.47&-21.86&-9.574&-8.532&-7.219&-18.71&-8.121& 34.05 s \\
     \hline
     $\nu_{-3}$&$\nu_{-2}$&$\nu_{-1}$&$\nu_{0}$&$\nu_{1}$&$\nu_{2}$&$\nu_{3}$&  \\
     \hline
     5.539 Hz& 7.039 Hz&8.539 Hz&10.039 Hz&11.539 Hz&13.039 Hz&14.539 Hz& \\
     \hline
\end{tabular}
    }
\end{center}
\end{table}

\bibliographystyle{naturemag}
\bibliography{supplementrefs}